\documentclass{article}
\usepackage{graphicx}
\usepackage[style=numeric, citestyle=numeric,sorting=none]{biblatex}
\usepackage{setspace}
\usepackage{amsfonts, amsmath, amsthm, amssymb}
\usepackage{booktabs}
\usepackage{xcolor}
\usepackage[left=1in,top=1in,right=1in,bottom=1in]{geometry}
\usepackage{multirow}
\usepackage{authblk}
\usepackage{hyperref}
\usepackage{cleveref}
\usepackage{algorithm}
\usepackage{algorithmic}
\usepackage{bbm}

\title{Investigating Targeting Strategies and Truncation in TMLE for the Average Treatment Effect under Practical Positivity Violations}

\author[1]{Yichen Xu}
\author[2]{Susan Gruber}
\author[1]{Mark J. van der Laan}

\affil[1]{Division of Biostatistics, University of California, Berkeley\\
\href{mailto:\{yichen\_xu,laan\}@berkeley.edu}{\{yichen\_xu,laan\}@berkeley.edu}}
\affil[2]{Targeted ML Solutions\\
\href{mailto:sgruber@targetedML.com}{sgruber@targetedML.com}}

\date{}

\begin{document}
\maketitle

\begin{abstract}

Estimating average treatment effects from observational data is challenging under practical violations of the positivity assumption. Targeted maximum likelihood estimators (TMLEs) are widely used because of their double robustness and efficiency, but their finite-sample performance can be sensitive to extreme inverse-probability factors in the targeting step. In standard TMLE implementations, these inverse-probability factors enter through the clever covariate used in the fluctuation regression; an alternative implementation removes these factors from the clever covariate and instead incorporates them as observation weights in the loss function, which we refer to as loss-weighted targeting. We conduct extensive simulation studies to examine how targeting strategy and truncation level affect TMLE performance under varying degrees of outcome-regression misspecification and practical positivity violations. We show that loss-weighted targeting can induce substantial systematic bias, whereas retaining the inverse-probability factor in the clever covariate, together with appropriate truncation, yields more stable performance. For clever-covariate-scaled targeting, insufficient truncation leads to inflated variance, indicating that the estimated propensity score must be bounded sufficiently far from zero to maintain estimator stability. We further find that the recommended settings $c = 5$ and $c = 6$ in the lower bound $c/(\sqrt{n}\ln n)$ provide robust fixed truncation rules across many settings, although the optimal truncation level varies with sample size. Building on these findings, we identify limitations of standard Lepski’s method for truncation selection in this setting and propose a Lepski-type adaptive truncation procedure with a safe brake mechanism designed to avoid unstable low-truncation regions. We also evaluate variance estimation methods for clever-covariate-scaled targeting and find that targeted bootstrap variance estimation provides a stable alternative across truncation levels.

\end{abstract}

\section{Introduction}
\label{sec:intro}

Observational studies are increasingly used to evaluate treatment and exposure effects in real-world settings, particularly as electronic health records, administrative databases, population-based surveys, and other large observational data sources have become more widely available \cite{fernainy24}. In observational studies, investigators do not randomly assign treatment or exposure but instead analyze nonrandomized data, such as electronic health records, administrative databases, population-based surveys, or single-arm trials augmented with external controls. Treatment decisions are typically driven by patient characteristics, clinical practice, or system-level factors rather than experimental allocation. Although the absence of randomization introduces the possibility of confounding, OS offer important advantages: they often reflect real-world conditions, enhance external validity, and may be more feasible, ethical, and cost-effective than RCTs, particularly in complex or large-scale settings. Without the benefit of randomization as in randomized controlled trials, observational studies require statistical adjustment for confounding. Broadly, such adjustment can proceed either through outcome regression methods—such as BART \cite{chipman10} used with the G-computation formula \cite{robins86}, which models the conditional mean of \(Y\) given treatment and covariates to estimate counterfactual outcomes—or through propensity score methods \cite{zhou20}, which model the treatment assignment mechanism to reweight or match subjects and thereby emulate randomization \cite{hernan16}. The propensity score represents the probability that an individual will be assigned to a particular treatment group, conditional on observed covariates. Methods that use the propensity score in estimation, such as inverse probability of treatment weighting (IPTW) \cite{chesnaye21} and targeted maximum likelihood estimation (TMLE) \cite{vdl11}, rely on the positivity assumption to ensure identification and stable estimation of causal effects \cite{zhu21}. This assumption, in theory, requires $0 < \text{PS} < 1$, meaning that the conditional probability of receiving any treatment cannot be zero or one within any confounder subgroup. A theoretical violation occurs when it is impossible for any subject in the subgroup to receive a certain treatment, and increasing the sample size cannot resolve this. In contrast, practical violation arises when a subject has a non-zero probability of treatment assignment, but the assignment does not occur in the observed data or is extremely rare. With sufficiently large sample sizes, practical violations can be mitigated, but with limited data, the estimated inverse PS, $\frac{1}{P(A=1|W)}$, can become extreme, leading to bias or inflated variance in the final estimators \cite{crump06, crump09}. To mitigate and diagnose the practical positivity issues in IPTW-based estimators, many works \cite{zhou20, joseph07, freedman08, cole08} truncate the extreme PS or filter out samples with extreme fitted PS. Other methods use overlap weights \cite{cheng22, matsouaka24} to upweight samples with higher overlapping scores. \cite{Petersen12} proposes to use bootstrapping as a method to identify positivity issues.

Targeted Maximum Likelihood Estimators (TMLE) \cite{vdl06, vdl11, vdl18, gruber12} are widely used for estimating causal parameters due to their desirable properties, including bias reduction, double robustness, and semiparametric efficiency \cite{smith23}. However, TMLE can be unstable under practical positivity violations. For example, \cite{qiu24} reports degraded performance when outcomes are rare and sample sizes are small. To improve stability, \cite{vdl23, vdl25} propose adaptive TMLE procedures that project the average treatment effect onto a working model defined by a relaxed highly adaptive lasso \cite{benseker16}, and \cite{gruber22} recommends truncating inverse probability weights at the level \( \frac{5}{\sqrt{n}\ln n} \) to control mean squared error. Regarding variance estimation, \cite{tran23} develops plug-in and targeted bootstrap approaches to address underestimation of the efficient influence function based variance estimator in longitudinal counterfactual mean settings. Despite these advances, systematic comparisons of loss-weighted and clever-covariate–scaled targeting across varying degrees of practical positivity violations remain limited. In addition, the stability of fixed truncation rules is not fully understood, data-adaptive truncation strategies remain underexplored, and the interaction between variance estimation and data-adaptive truncation procedures, such as Lepski-type selection, has received limited attention for the point-treatment average treatment effect.

In this article, we focus on continuous outcomes, following \cite{gruber10}, as they present more nuanced challenges than binary outcomes. We conduct extensive simulation studies across multiple scenarios, including varying degrees of outcome regression misspecification (from highly misspecified to nearly correctly specified), the severity of practical positivity violations, a comparison of \emph{loss-weighted targeting} and \emph{clever-covariate-scaled targeting} in the TMLE update step \cite{gruber13}, and different truncation levels $c$ in the inverse-weight bound $\frac{c}{\sqrt{n}\ln n}$. Here, \emph{loss-weighted targeting} (\texttt{gWt}) refers to a TMLE fluctuation that incorporates inverse-probability factors directly into the loss function, whereas \emph{clever-covariate-scaled targeting} (\texttt{gH}) keeps the loss unweighted and instead incorporates the inverse-probability factors through the clever covariate in the fluctuation submodel. We demonstrate that \emph{loss-weighted targeting} can induce substantial systematic bias under practical positivity violations, whereas retaining the inverse-probability factor in the clever covariate and applying appropriate truncation yields more stable performance. For \emph{clever-covariate-scaled targeting}, minimal truncation leads to increased variance, indicating that a sufficiently large truncation level is necessary to maintain estimator stability. Building on these findings, we identify limitations of standard Lepski’s method for truncation selection in this context and propose a Lepski-type adaptive truncation mechanism with a brake rule that preserves stability when the Lepski selection is unreliable, enabling robust data-adaptive truncation in practice.

The remainder of the paper is organized as follows. In \Cref{sec:tmle}, we review the TMLE framework and describe the two targeting strategies examined in this work. In \Cref{sec:lepksi}, we introduce the Lepski-with-brake procedure for data-adaptive selection of the truncation level. In \Cref{sec:simulation}, we detail the simulation design, present results from our extensive simulation study, and benchmark various variance estimation methods. Additional investigations comparing linear and logistic links, as well as the performance of \texttt{gwt} and \texttt{gh} in settings without positivity violations, are provided in the Appendix.

\section{Targeting strategies and variance estimation}
\label{sec:tmle}

\subsection*{TMLE construction}

Targeted maximum likelihood estimation (TMLE) is a two-stage, loss-based procedure that yields a doubly robust, semiparametrically efficient substitution estimator of a prespecified target parameter. An initial data‐adaptive estimator of the relevant portion of the data-generating distribution is first obtained without regard to the target parameter. This fit is then updated through a low-dimensional parametric fluctuation model whose score spans the efficient influence function of the target parameter. The fluctuation parameter is estimated by minimizing a targeted loss, yielding an updated estimator that approximately solves the efficient influence curve estimating equation. This targeting step removes the leading (first-order) bias of the plug-in estimator while respecting global model constraints, resulting in an estimator with improved finite-sample bias–variance tradeoff and valid asymptotic inference even when machine-learning methods are used for nuisance estimation.

Concretely, let the observed data be $O = (W, A, Y)$, where $W$ denotes the covariates, $A \in \{0,1\}$ is a binary treatment indicator, and $Y$ is the outcome. The parameter of interest is the average treatment effect (ATE).
\[
\psi_0=\mathbb{E}\!\left[Q_0(1,W)-Q_0(0,W)\right],
\]
where $Q_0(a,W)=\mathbb{E}(Y\mid A=a,W)$ and $g_0(W)=P(A=1\mid W)$.  
TMLE begins with estimation of the outcome regression $\bar Q_t(A,W)\approx Q_0(A,W)$ and the propensity score $\hat g(W)\approx g_0(W)$.  
The targeting step updates $\bar Q_t$ through a fluctuation model chosen to reduce bias for the target parameter. In our implementation, the two counterfactual means under treatment and control are targeted separately, using the arm-specific clever covariates
\[
H_{a=1}(A,W)=\frac{A}{\hat g(W)},
\qquad
H_{a=0}(A,W)=\frac{1-A}{1-\hat g(W)}.
\]
Accordingly, separate fluctuation parameters $\epsilon_1$ and $\epsilon_0$ are used to update the outcome regression for the treated and control arms, respectively, and we consider two alternative targeting strategies for carrying out this update.

\textbf{Loss-weighted targeting \cite{gruber13}.}
The logit fluctuation applies inverse probability weights directly to the loss:
\begin{align*}
\mathcal{L}_{\texttt{gWt}}(\bar Q(\epsilon_{t+1}))
&=
-\sum_{i=1}^n
\Bigg[
\frac{A_i}{\hat g(W_i)}
\Big\{
\tilde Y_i \log p_{1i}
+
(1-\tilde Y_i)\log(1-p_{1i})
\Big\}
\\
&\qquad\qquad+
\frac{1-A_i}{1-\hat g(W_i)}
\Big\{
\tilde Y_i \log p_{0i}
+
(1-\tilde Y_i)\log(1-p_{0i})
\Big\}
\Bigg],
\end{align*}
where
\begin{align*}
p_{1i}
&=
\frac{
1
}{
1+
\exp\!\left(
-
\log\frac{\bar Q_{t,i}}{1-\bar Q_{t,i}}
+
\epsilon_1
\right)
},
\\
p_{0i}
&=
\frac{
1
}{
1+
\exp\!\left(
-
\log\frac{\bar Q_{t,i}}{1-\bar Q_{t,i}}
+
\epsilon_0
\right)
}.
\end{align*}
This formulation corresponds to a weighted likelihood update with separate fluctuation parameters for the treated and control groups, and yields a solution to the associated score equation.

\textbf{Clever-covariate-scaled targeting \cite{vdl11, vdl18}.}
The loss remains unweighted and inverse probability factors enter through the clever covariate term:
\begin{align*}
\mathcal{L}_{\texttt{gH}}(\bar Q(\epsilon_{t+1}))
&=
-
\sum_{i=1}^n
\Big[
\tilde Y_i \log p_i
+
(1-\tilde Y_i)\log(1-p_i)
\Big],
\end{align*}
with
\begin{align*}
p_i
=
\frac{
1
}{
1+
\exp\!\left(
-
\log\frac{\bar Q_{t,i}}{1-\bar Q_{t,i}}
+
\epsilon_1\frac{A_i}{\hat g(W_i)}
+
\epsilon_0\frac{1-A_i}{1-\hat g(W_i)}
\right)
}.
\end{align*}
After estimating $(\epsilon_1,\epsilon_0)$, the targeted regression $\bar Q_t^*(a,W)$ is obtained by evaluating the potential outcome under each counterfactual treatment assignment using the updated regression model. The ATE is then computed as a plug-in estimator.
\[
\hat\psi
=
\frac1n
\sum_{i=1}^n
\Big[
\bar Q_t^*(1,W_i)-\bar Q_t^*(0,W_i)
\Big].
\]

To stabilize targeting under practical positivity violations, let $\hat g_1(W)=\hat P(A=1\mid W)$ and $\hat g_0(W)=\hat P(A=0\mid W)=1-\hat g_1(W)$ denote the estimated treatment and control assignment probabilities, respectively. We truncate these estimated propensity scores according to
\[
\hat g_1(W)\leftarrow \max\{\hat g_1(W),b\},
\qquad
\hat g_0(W)\leftarrow \max\{\hat g_0(W),b\},
\qquad
b=\frac{c}{\sqrt n \ln n}.
\]

\subsection*{Variance estimation}

After obtaining the estimate, one should estimate its variance for construction of Wald-type confidence intervals. The sample variance of the efficient influence function is an asymptotically valid variance estimator for TMLEs:
\begin{align*}
\widehat{\mathrm{Var}}(\hat\psi)
=
\frac{1}{n}
\mathbb{P}_n\!\left[D^*(O)^2\right]
=
\frac{1}{n^2}
\sum_{i=1}^n
\big(D^*(O_i)\big)^2,
\end{align*}
where in implementation, we use
\begin{align*}
D^*(O_i)
=
\frac{A_i}{\hat g_1(W_i)}\big(Y_i-\hat Q^*(1,W_i)\big)
-
\frac{1-A_i}{\hat g_0(W_i)}\big(Y_i-\hat Q^*(0,W_i)\big)
+
\hat Q^*(1,W_i)-\hat Q^*(0,W_i)
-
\hat\psi.
\end{align*}
Under asymptotic linearity of TMLE,
\begin{align*}
\hat\psi-\psi_0
=
\frac{1}{n}\sum_{i=1}^n D^*(O_i)
+
o_p(n^{-1/2}).
\end{align*}
Thus, this estimator consistently estimates the asymptotic variance of $\hat\psi$ under standard regularity conditions. However, the EIF-based variance estimator may underestimate variability in the presence of practical positivity violations. When certain $(A,W)$ combinations are rare, the corresponding inverse probability weights, $\frac{1}{\hat g_1(W)}$ or $\frac{1}{\hat g_0(W)}$, can be extremely large but infrequently realized in finite samples, leading to downward bias in the empirical variance estimate.

A plug-in variance estimator \cite{tran23} is obtained by decomposing the variance of the efficient influence function into a residual term and a regression term. For continuous outcomes, let $\hat\sigma^2(a,W)$ denote an estimate of $\mathrm{Var}(Y \mid A=a, W)$. Then
\begin{align*}
\widehat{\mathrm{Var}}_{\text{plug-in}}(\hat\psi)
&=
\frac{1}{n}
\sum_{i=1}^{n}
\left[
\frac{\hat\sigma^2(1,W_i)}{\hat g(1|W_i)}
+
\frac{\hat\sigma^2(0,W_i)}{\hat g(0|W_i)}
\right]
+
\frac{1}{n}
\sum_{i=1}^n
\left[
\hat Q(1,W_i)-\hat Q(0,W_i)-\hat\psi
\right]^2.
\end{align*}
Compared with the EIF-based estimator, the inverse probability factors in the plug-in estimator are multiplied by the estimated conditional variance. This can reduce the effect of extreme weights and help avoid variance underestimation under practical positivity violations.

As a further alternative, we consider a targeted bootstrap variance estimator \cite{tran23}. In this procedure, the initial nuisance estimators $\hat g$ and $\hat Q$ are fit once on the original dataset and kept fixed. For $b=1,\dots,B$, we draw a bootstrap sample with replacement and use the previously fitted $\hat g(W)$ and $\hat Q(a,W)$ for the resampled observations. Within each bootstrap sample, we re-estimate the fluctuation parameter $\epsilon^{*(b)}$, yielding an updated targeted estimator $\hat\psi^{*(b)}$. The targeted bootstrap variance estimator is defined as
\begin{align*}
\widehat{\mathrm{Var}}_{\text{TB}}(\hat\psi)
=
\frac{1}{B-1}
\sum_{b=1}^B
\left(
\hat\psi^{*(b)}-\bar\psi^*
\right)^2,
\end{align*}
where
$\bar\psi^*
=
\frac{1}{B}
\sum_{b=1}^B
\hat\psi^{*(b)}$.

By re-estimating only the targeting fluctuation step while holding the initial nuisance fits fixed, this procedure isolates variability induced by the inverse-weighting and update step without incurring the computational cost of repeated nuisance re-fitting. In addition, one may base inference directly on bootstrap quantiles rather than the empirical variance when a more outlier-robust summary of the bootstrap distribution is desired.

\section{Adaptive truncation via a Lepski with brake procedure}

\label{sec:lepksi}

The impact of practical positivity violations on estimator performance may vary substantially across data sets and across levels of outcome model misspecification. While a fixed truncation rule such as $c=5$ provides a robust and convenient default, it represents a heuristic compromise that may be suboptimal for a given sample: weak truncation can induce extreme inverse–probability weights and unstable estimation with inflated variance, whereas overly aggressive truncation can increase bias by distorting the effective target population. To avoid relying on a single prespecified truncation level, we therefore develop a data-adaptive strategy that selects the degree of truncation based on observed estimator stability. This motivates the use of a Lepski‐type \cite{lepski91, lepski92} procedure, which compares estimator displacement across truncation levels to their associated statistical uncertainty and advances only when changes cannot be explained by sampling variability alone.

Let $c_1 < c_2 < \cdots < c_K$ denote an ascending sequence of truncation constants, where larger values of $c$ correspond to stronger truncation of inverse propensity weights. For each $c_k$, we compute a TMLE point estimate $\hat{\psi}(c_k)$ together with an efficient influence function (EIF)-based variance estimate $\hat V(c_k)$, yielding the Wald confidence interval
\[
I(c_k)
=
\bigl[
\hat{\psi}(c_k) \pm 1.96 \sqrt{\hat V(c_k)}
\bigr].
\]

The truncation level is selected adaptively by moving sequentially along the grid from stronger truncation toward weaker truncation and comparing adjacent estimators. In other words, starting from a large truncation constant, we move to the next smaller level of truncation only when the resulting change in the point estimate is large relative to the accompanying increase in uncertainty, suggesting a worthwhile reduction in truncation bias. Specifically, movement from $c_k$ to $c_{k-1}$ is allowed if and only if
\[
\Bigl(
\hat{\psi}(c_{k-1}) < \hat{\psi}(c_k)
\ \text{and}\
\sup I(c_{k-1}) \le \sup I(c_k)
\Bigr)
\]
or
\[
\Bigl(
\hat{\psi}(c_{k-1}) > \hat{\psi}(c_k)
\ \text{and}\
\inf I(c_{k-1}) \ge \inf I(c_k)
\Bigr).
\]
These conditions require that the entire confidence interval at the next truncation level lies beyond that of the current level in the same direction as the point estimate movement. Thus, the procedure advances only when the shift of the estimator cannot be explained by the widening of statistical uncertainty, reflecting the classical Lepski logic of comparing estimator displacement to confidence-interval width. When neither inequality holds, the confidence intervals overlap too substantially to justify further movement along the truncation path. In this case, the algorithm stops and selects $\hat c = c_k$.

Consider the sequence of targeted estimators $\{\hat\psi(c): c\in\mathcal C\}$ obtained along a truncation path, and let $\psi_0$ denote the true target parameter. 
Under standard asymptotic linearity,
\begin{equation}
\hat\psi(c) - \psi_0
=
(P_n-P)D_c + r_n(c),
\label{eq:linrep_clean}
\end{equation}
where $D_c$ is the efficient influence function at truncation level $c$ and $|r_n(c)| = o_p(n^{-1/2})$ asymptotically. 
Consequently, each estimator fluctuates around $\psi_0$ at the canonical $n^{-1/2}$ scale in large samples.

Our objective is to construct a region that plausibly contains $\psi_0$ and to provide a natural brake that avoids selecting a truncation level with too small a value of $c$. Because $\psi_0$ is unknown, we center this region at a stable anchor estimator, chosen as the most-truncated estimate $\hat\psi(c_{\max})$, which satisfies
\[
|\hat\psi(c_{\max})-\psi_0| = O_p(n^{-1/2}).
\]

In finite samples, practical positivity violations may induce additional variability along the truncation path. 
When estimated propensity scores approach zero, the efficient influence function can become highly variable, leading to realized deviations of $\hat\psi(c)$ from $\psi_0$ that exceed the nominal $n^{-1/2}$ scale despite asymptotic linearity. 
To accommodate such instability without prematurely excluding reasonable truncation levels, we enlarge the localization radius by a slowly varying factor and define
\begin{equation}
\mathcal E_n
=
\left\{
\theta:
|\theta-\hat\psi(c_{\max})|
\le
z_n\,\widehat{\mathrm{SE}}(c_{\max})
\right\},
\label{eq:env_clean}
\end{equation}
where $\widehat{\mathrm{SE}}(c_{\max})$ estimates the standard error of the estimator at $c_{\max}$ and $z_n$ grows slowly with $n$, for example $z_n \propto \sqrt{\log n}$. Since $\widehat{\mathrm{SE}}(c_{\max})$ is typically of order $n^{-1/2}$, the resulting tolerance radius satisfies
\[
z_n\,\widehat{\mathrm{SE}}(c_{\max})
\;\asymp\;
\sqrt{\frac{\log n}{n}},
\]
which vanishes asymptotically while remaining slightly larger than the classical $n^{-1/2}$ rate. This inflation is not intended to characterize a sharper asymptotic rate; rather, it provides a conservative margin that mitigates finite-sample irregularities arising from near-positivity violations.

Truncation levels for which $\hat\psi(c)\in\mathcal E_n$ are regarded as compatible with the anchor region. 
Adaptive truncation is then implemented by applying a Lepski selection rule restricted to this localized set. 
The resulting procedure acts as a safeguard that preserves stability under finite-sample instability while maintaining a near–root-$n$ contraction of the admissible region as the sample size increases.

\section{Simulation}

\label{sec:simulation}

\subsection{Settings}

We consider a factorial simulation design that varies both the severity of practical positivity violations and the degree of outcome regression misspecification. For each scenario, we generate independent observations
$O_i = (W_i, A_i, Y_i)$ for $i = 1,\dots,n$, where $A_i \in \{0,1\}$ denotes treatment assignment, $W_i = (W_{i1},W_{i2},W_{i3},W_{i4})$ is a vector of baseline covariates, and $Y_i$ is a continuous outcome. The data generating process is defined in \eqref{eq:dgp}.

Under both the high and moderate misspecification settings, outcomes are generated from the same nonlinear outcome model. The difference lies in the working outcome regression used for estimation: under high misspecification, the fitted outcome regression omits \(W_1\), whereas under moderate misspecification all covariates are included. Under the nearly correct specification setting, outcomes are generated from the simpler model above and all covariates are included in the fitted outcome regression. Thus, the three settings induce different levels of misspecification relative to the Gaussian linear working model used for outcome regression, while the propensity score is estimated using a correctly specified logistic regression with all covariates.

\begin{equation}
\label{eq:dgp}
\begin{array}{rcl}
W_{ij} &\sim& \mathrm{Uniform}(-1,1), \qquad j=1,2,3,4, \\[1.2ex]
\eta_i &=& \kappa_{\mathrm{pos}}\bigl(1.5\,W_{i1} + 2\,W_{i2} - W_{i3} - 2.5\,W_{i4}\bigr), \\[1.2ex]
\pi_{0,i} &=& \mathrm{expit}(\eta_i), \\[1.2ex]
A_i &\sim& \mathrm{Bernoulli}(\pi_{0,i}), \\[2ex]
\mu_{0,i}
&=&
\begin{cases}
W_{i1} + |W_{i2}| + W_{i3} + |W_{i4}|, & \text{(high and moderate misspecification)},\\[1.2ex]
W_{i1} + W_{i2} + W_{i3} + W_{i4}, & \text{(nearly correct specification)},
\end{cases}
\\[2.2ex]
\tau_i
&=&
\begin{cases}
3 + 2\,W_{i1} - 0.5|W_{i2}| + 0.5\,W_{i3} - 2|W_{i4}|, & \text{(high and moderate misspecification)},\\[1.2ex]
2 - 0.5|W_{i1}|, & \text{(nearly correct specification)},
\end{cases}
\\[2.2ex]
Y_i &\sim& \mathcal{N}\!\Big(\mu_{0,i} + A_i \tau_i,\ 0.5^2 \Big).
\end{array}
\end{equation}

For each scenario, we generate \(500\) independent Monte Carlo data sets with sample size \(n \in \{500, 1{,}000, 2{,}000\}\). The propensity score scaling constant \(\kappa_{\mathrm{pos}} \in \{1,2,3\}\) controls overlap at the data-generating stage. When \(\kappa_{\mathrm{pos}}=3\), the simulated propensity scores range from approximately \(1.7\times10^{-7}\) to values arbitrarily close to \(1\); when \(\kappa_{\mathrm{pos}}=2\), the minimum propensity score is approximately \(3.1\times10^{-5}\); and when \(\kappa_{\mathrm{pos}}=1\), the propensity scores range from approximately \(0.0055\) to \(0.998\). Truncation is imposed at the estimation stage. For \(n=1{,}000\), the truncation threshold \(c/(\sqrt{n}\ln n)\) equals approximately \(0.0046\) for \(c=1\), \(0.0092\) for \(c=2\), and \(0.0229\) for the recommended value \(c=5\). Thus, when \(\kappa_{\mathrm{pos}}\in\{2,3\}\), the minimum propensity scores fall below even the weakest of these thresholds and truncation is activated. When \(\kappa_{\mathrm{pos}}=1\), the minimum propensity score is approximately \(0.0055\), so truncation is minimal for \(c=1\) but occurs for larger values of \(c\), including the recommended setting \(c=5\).

Each simulated data set is analyzed using two TMLE targeting strategies: (i) loss-weighted targeting and (ii) clever-covariate-scaled targeting. For both targeting strategies, truncation of the inverse weights is applied using bounds of the form $c/(\sqrt{n}\ln n)$ over a grid of constants $c \in \{1,\ldots,10\}$. In addition, for the clever-covariate-scaled targeting strategy, we evaluate the Lepski-type adaptive truncation procedure with a brake, which selects the truncation level data-adaptively while enforcing stability when the Lepski criterion becomes unreliable.

Estimator performance is evaluated in terms of bias, variance, mean squared error, the ratio of bias to standard error, and confidence interval coverage, based on 500 Monte Carlo replications for each simulation scenario. Using the same simulated datasets, we additionally benchmark three variance estimators for the clever-covariate-scaled targeting strategy: (i) the efficient influence function (EIF)-based estimator, (ii) the plug-in estimator, and (iii) the targeted bootstrap (TB) estimator. Because the Lepski stopping and braking rules depend on variance estimates, we also compare the resulting Lepski-based procedures under each variance choice. Specifically, we denote the corresponding methods as EIF-brake (EIFb), Monte Carlo-brake (MCb), and targeted bootstrap-brake (TBb).

\subsection{Clever-covariate-scaled targeting is more effective than loss-weighted targeting under practical positivity violations}

\begin{figure}[t!]
  \centering
  \includegraphics[width=0.95\textwidth]{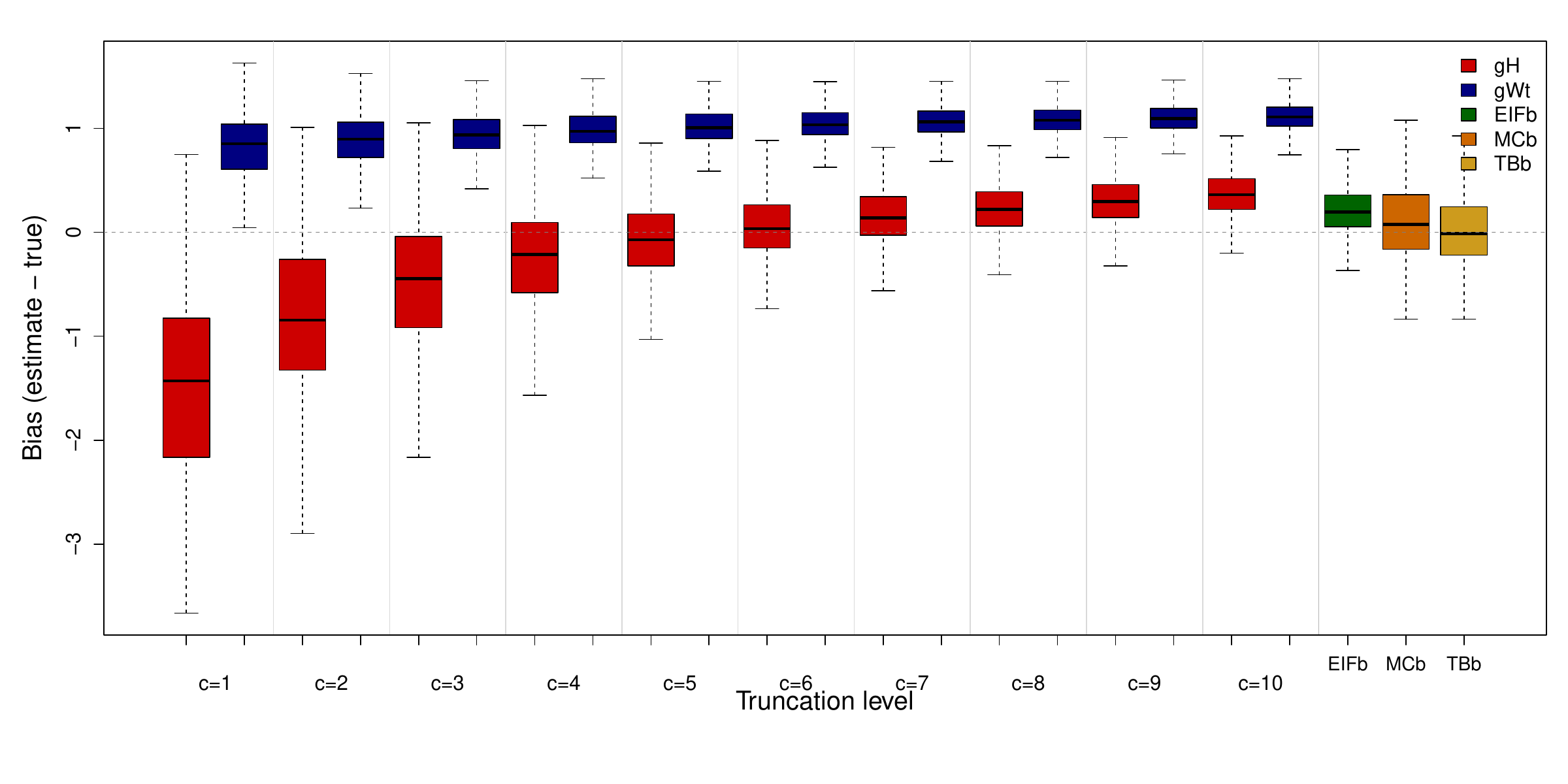}
  \caption{EIFb/MCb/TBb adaptive truncation (n=1000): high misspecification, severe positivity.}
  \label{fig:n1000_mu2pos3}
\end{figure}

Even though, in the absence of practical positivity violations, loss-weighted targeting and clever-covariate–scaled targeting exhibit nearly identical performance (see \Cref{fig:mse-ratio-fluct}), their behavior diverges systematically when practical positivity violations occur. This pattern holds consistently across sample sizes \(n \in \{500,1000,2000\}\), across moderate and high outcome model misspecification, and across stronger positivity violations with \(\kappa_{\mathrm{pos}} \in \{2,3\}\). To illustrate this phenomenon, we focus on a representative scenario with high outcome misspecification and severe positivity violations (\(\kappa_{\mathrm{pos}} = 3\)) at sample size \(n=1000\), shown in \Cref{fig:n1000_mu2pos3} and \Cref{tab:n1000_mu2pos3_detail}.

For the loss-weighted targeting approach (gWt), increasing the truncation level \(c\) primarily reduces the variance but does not correct the large systematic bias. For example, the variance decreases from \(0.112\) at \(c=1\) to \(0.019\) at \(c=10\), while the bias increases from \(0.804\) to \(1.114\). As a result, the mean squared error remains large across the truncation path (e.g., \(0.758\) at \(c=1\) and \(1.259\) at \(c=10\)). The bias-to-standard-error ratio is extremely large throughout, increasing from \(2.40\) at \(c=1\) to over \(8.07\) at \(c=10\), indicating severe bias relative to statistical uncertainty. Consequently, the coverage probability collapses rapidly, dropping from \(0.30\) at \(c=1\) to essentially zero for \(c \ge 6\). This shows that gWt fails to achieve reasonable inference under practical positivity violations.

In contrast, the clever-covariate–scaled targeting approach (gH) exhibits a markedly different pattern. When truncation is weak, both bias and variance are large; for instance, at \(c=1\) the bias is \(-1.43\), the variance is \(0.846\), and the MSE is \(2.90\). As the truncation level increases, both bias and variance decrease substantially. The bias shrinks from \(-1.43\) at \(c=1\) to \(-0.077\) at \(c=5\) and \(0.049\) at \(c=6\), while the variance decreases from \(0.846\) to \(0.147\) and \(0.106\), respectively. This leads to a dramatic reduction in MSE, reaching \(0.153\) at \(c=5\) and \(0.108\) at \(c=6\). Correspondingly, the bias-to-standard-error ratios become small (\(0.20\) at \(c=5\) and \(0.15\) at \(c=6\)), indicating that the remaining bias is negligible relative to sampling variability. The resulting coverage probabilities are \(0.946\) at \(c=5\) and \(0.942\) at \(c=6\), both close to the nominal \(95\%\) level. For larger truncation levels, truncation-induced bias begins to dominate: the bias increases again (e.g., \(0.302\) at \(c=9\) and \(0.363\) at \(c=10\)), while the variance continues to decrease, leading to increasing bias-to-SE ratios and deteriorating coverage.

The adaptive truncation procedures select truncation levels within this stable intermediate region. For example, the targeted bootstrap brake (TBb) achieves bias \(0.027\), variance \(0.113\), and MSE \(0.114\), with a small bias-to-SE ratio of \(0.080\) and coverage \(0.956\). Similarly, the Monte Carlo brake (MCb) yields bias \(0.104\), variance \(0.119\), and coverage \(0.948\). These results suggest that the adaptive truncation procedures can select truncation levels in a stable intermediate region in this representative setting, avoiding the most severe high-variance behavior associated with weak truncation while not moving as far into the strongly truncated, bias-dominated regime.

\begin{table}[t!]
\centering
\scriptsize
\begin{tabular}{llrrrrr}
\toprule
Method & Setting & Bias & SE & $|\mathrm{Bias}|/\mathrm{SE}$ & MSE & Coverage \\
\midrule
\multirow{10}{*}{gH}
& $c=1$  & -1.432 & 0.920 & 1.557 & 2.897 & 0.630 \\
& $c=2$  & -0.838 & 0.790 & 1.060 & 1.325 & 0.814 \\
& $c=3$  & -0.481 & 0.606 & 0.793 & 0.598 & 0.866 \\
& $c=4$  & -0.245 & 0.472 & 0.518 & 0.283 & 0.908 \\
& $c=5$  & -0.077 & 0.384 & 0.202 & 0.153 & 0.946 \\
& $c=6$  &  0.049 & 0.326 & 0.149 & 0.108 & 0.942 \\
& $c=7$  &  0.149 & 0.285 & 0.523 & 0.104 & 0.920 \\
& $c=8$  &  0.232 & 0.255 & 0.909 & 0.119 & 0.838 \\
& $c=9$  &  0.302 & 0.232 & 1.301 & 0.145 & 0.742 \\
& $c=10$ &  0.363 & 0.215 & 1.688 & 0.178 & 0.614 \\
\midrule
\multirow{10}{*}{gWt}
& $c=1$  & 0.804 & 0.335 & 2.398 & 0.758 & 0.300 \\
& $c=2$  & 0.882 & 0.251 & 3.512 & 0.841 & 0.068 \\
& $c=3$  & 0.936 & 0.212 & 4.409 & 0.922 & 0.008 \\
& $c=4$  & 0.977 & 0.190 & 5.141 & 0.990 & 0.002 \\
& $c=5$  & 1.009 & 0.174 & 5.791 & 1.048 & 0.002 \\
& $c=6$  & 1.036 & 0.163 & 6.367 & 1.099 & 0.000 \\
& $c=7$  & 1.059 & 0.154 & 6.876 & 1.145 & 0.000 \\
& $c=8$  & 1.079 & 0.147 & 7.325 & 1.186 & 0.000 \\
& $c=9$  & 1.097 & 0.142 & 7.715 & 1.224 & 0.000 \\
& $c=10$ & 1.114 & 0.138 & 8.066 & 1.259 & 0.000 \\
\midrule
{EIFb} & Adaptive & 0.210 & 0.229 & 0.918 & 0.097 & 0.848 \\
{MCb}  & Adaptive & 0.104 & 0.345 & 0.301 & 0.130 & 0.948 \\
{TBb}  & Adaptive & 0.027 & 0.336 & 0.080 & 0.114 & 0.956 \\
\bottomrule
\end{tabular}
\caption{Detailed simulation results for the representative setting shown in \Cref{fig:n1000_mu2pos3}: \(n=1000\), high outcome-regression misspecification, and severe practical positivity violations \((\kappa_{\mathrm{pos}}=3)\).}
\label{tab:n1000_mu2pos3_detail}
\end{table}

\subsection{Lepski adaptive truncation level selection with brake}

We evaluate the adaptive truncation procedures by aggregating results across all simulation scenarios. Performance is summarized using the mean, median, and worst-case MSE across scenarios, together with coverage accuracy measured by the coverage error, defined as the positive part of the difference between the nominal level $0.95$ and the empirical coverage so that only undercoverage is penalized. For each method we report the mean, median, and worst coverage error across scenarios. These summaries are computed over the full collection of simulation settings combining three positivity levels ($\kappa_{\mathrm{pos}}\in\{1,2,3\}$) and three outcome misspecification levels across sample size $500$, $1000$, $2000$. The aggregated results are reported in Table~\ref{tab:lepski_summary}.

A first observation from Table~\ref{tab:lepski_summary} is that extreme truncation levels are not robust across sample sizes. Very small truncation levels ($c=1$--$3$) consistently lead to extremely large MSE. For example, when $n=500$, $c=1$ yields mean MSE $0.6730$ and worst MSE $3.1984$, and even when $n=2000$ the mean MSE remains very large ($0.5092$). At the other extreme, very large truncation levels ($c=8$--$10$) produce substantial coverage deterioration in smaller samples: when $n=500$, $c=10$ results in mean coverage error $0.1582$ and worst coverage error $0.708$. These results indicate that both overly small and overly large truncation levels can lead to unstable inference.

Second, the table shows a clear trend that the optimal truncation level shifts with the sample size. When $n=500$, moderate truncation performs best: for example, $c=5$ achieves the smallest mean MSE ($0.0557$), while $c=4$ also performs well with mean MSE $0.0726$. However, the performance of these choices deteriorates as the sample size increases. For instance, $c=4$ performs reasonably at $n=500$ but becomes clearly suboptimal at $n=1000$ (mean MSE $0.0763$) and $n=2000$ (mean MSE $0.0971$). Similarly, although $c=5$ is optimal at $n=500$, its performance worsens at $n=1000$ (mean MSE $0.0474$) and becomes particularly poor at $n=2000$ (mean MSE $0.0605$). In contrast, slightly larger truncation levels become preferable as the sample size grows: $c=6$ improves performance at $n=1000$ (mean MSE $0.0360$), and when $n=2000$ even larger truncation levels ($c=7$--$10$) achieve the smallest MSE together with smallest coverage error. Overall, the results indicate that the optimal truncation level increases with the sample size in these experiments.

Finally, the adaptive Lepski procedures with brake provide a data-driven strategy that avoids relying on a single fixed truncation level. In particular, the TB-based brake (TBb) performs competitively relative to recommended fixed choices such as $c=5$ and $c=6$. When $n=500$, TBb attains mean MSE $0.0610$ with mean coverage error $0.0176$, remaining close to the performance of $c=5$ while improving coverage relative to $c=6$ (mean coverage error $0.0242$). When $n=1000$, TBb substantially improves worst-case MSE compared with $c=5$ (0.1137 versus 0.1531) while maintaining similar coverage accuracy. When $n=2000$, TBb yields both smaller mean MSE and smaller coverage error than $c=5$ (mean MSE $0.0428$ versus $0.0605$, mean coverage error $0.0169$ versus $0.0293$). These results suggest that the adaptive procedure provides a competitive alternative to fixed truncation rules, although it is not uniformly superior to recommended fixed settings such as $c=5$ or $c=6$.

\begin{table}[t!]
\centering
\scriptsize

\begin{tabular}{lcccccc}
\hline
Method & mean MSE & median MSE & worst MSE & mean cov err & median cov err & worst cov err \\
\hline

\multicolumn{7}{c}{\textbf{n = 500}} \\
\hline
{EIFb} & 0.068 & 0.039 & 0.256 & 0.052 & 0.004 & 0.338 \\
{MCb}  & 0.061 & 0.034 & 0.206 & 0.022 & 0.014 & 0.106 \\
{TBb}  & 0.061 & 0.032 & 0.189 & 0.018 & 0.012 & 0.060 \\
gH\_c5 & 0.056 & 0.026 & 0.172 & 0.006 & 0.004 & 0.026 \\
gH\_c6 & 0.056 & 0.024 & 0.202 & 0.024 & 0.012 & 0.112 \\
\hline
gH\_c1 & 0.673 & 0.198 & 3.198 & 0.081 & 0.012 & 0.346 \\
gH\_c2 & 0.272 & 0.077 & 1.126 & 0.034 & 0.010 & 0.126 \\
gH\_c3 & 0.126 & 0.046 & 0.422 & 0.014 & 0.014 & 0.040 \\
gH\_c4 & 0.073 & 0.033 & 0.220 & 0.005 & 0.002 & 0.022 \\
\hline
gH\_c7 & 0.062 & 0.024 & 0.255 & 0.047 & 0.014 & 0.244 \\
gH\_c8 & 0.071 & 0.025 & 0.317 & 0.081 & 0.026 & 0.428 \\
gH\_c9 & 0.081 & 0.028 & 0.382 & 0.126 & 0.044 & 0.606 \\
gH\_c10 & 0.092 & 0.031 & 0.446 & 0.158 & 0.058 & 0.708 \\
\hline

\multicolumn{7}{c}{\textbf{n = 1000}} \\
\hline
{EIFb} & 0.036 & 0.029 & 0.097 & 0.021 & 0.010 & 0.102 \\
{MCb}  & 0.043 & 0.025 & 0.130 & 0.011 & 0.014 & 0.022 \\
{TBb}  & 0.043 & 0.022 & 0.114 & 0.011 & 0.008 & 0.030 \\
gH\_c5 & 0.047 & 0.022 & 0.153 & 0.009 & 0.006 & 0.022 \\
gH\_c6 & 0.036 & 0.018 & 0.108 & 0.008 & 0.008 & 0.018 \\
\hline
gH\_c1 & 0.573 & 0.161 & 2.897 & 0.075 & 0.004 & 0.320 \\
gH\_c2 & 0.280 & 0.063 & 1.325 & 0.039 & 0.008 & 0.156 \\
gH\_c3 & 0.141 & 0.038 & 0.598 & 0.028 & 0.014 & 0.088 \\
gH\_c4 & 0.076 & 0.028 & 0.283 & 0.017 & 0.008 & 0.050 \\
\hline
gH\_c7 & 0.032 & 0.015 & 0.104 & 0.013 & 0.010 & 0.030 \\
gH\_c8 & 0.033 & 0.013 & 0.119 & 0.028 & 0.014 & 0.112 \\
gH\_c9 & 0.036 & 0.012 & 0.145 & 0.046 & 0.022 & 0.208 \\
gH\_c10 & 0.041 & 0.012 & 0.178 & 0.071 & 0.028 & 0.336 \\
\hline

\multicolumn{7}{c}{\textbf{n = 2000}} \\
\hline
{EIFb} & 0.025 & 0.021 & 0.055 & 0.015 & 0.002 & 0.072 \\
{MCb}  & 0.039 & 0.018 & 0.147 & 0.016 & 0.012 & 0.056 \\
{TBb}  & 0.043 & 0.015 & 0.150 & 0.017 & 0.008 & 0.088 \\
gH\_c5 & 0.061 & 0.020 & 0.277 & 0.029 & 0.000 & 0.120 \\
gH\_c6 & 0.040 & 0.016 & 0.162 & 0.020 & 0.008 & 0.074 \\
\hline
gH\_c1 & 0.509 & 0.113 & 2.893 & 0.077 & 0.004 & 0.362 \\
gH\_c2 & 0.291 & 0.043 & 1.619 & 0.056 & 0.006 & 0.220 \\
gH\_c3 & 0.165 & 0.029 & 0.878 & 0.043 & 0.008 & 0.174 \\
gH\_c4 & 0.097 & 0.024 & 0.486 & 0.036 & 0.002 & 0.148 \\
\hline
gH\_c7 & 0.028 & 0.013 & 0.100 & 0.012 & 0.006 & 0.048 \\
gH\_c8 & 0.021 & 0.011 & 0.068 & 0.008 & 0.008 & 0.018 \\
gH\_c9 & 0.018 & 0.009 & 0.055 & 0.008 & 0.002 & 0.032 \\
gH\_c10 & 0.016 & 0.008 & 0.052 & 0.010 & 0.008 & 0.034 \\
\hline

\end{tabular}

\caption{Simulation performance across truncation levels and adaptive selectors.
Coverage error is defined as $\max(0,\,0.95 - \text{Coverage})$.}
\label{tab:lepski_summary}
\end{table}

\subsection{Variance estimation}

\begin{figure}[t]
    \centering
    \includegraphics[width=0.95\textwidth]{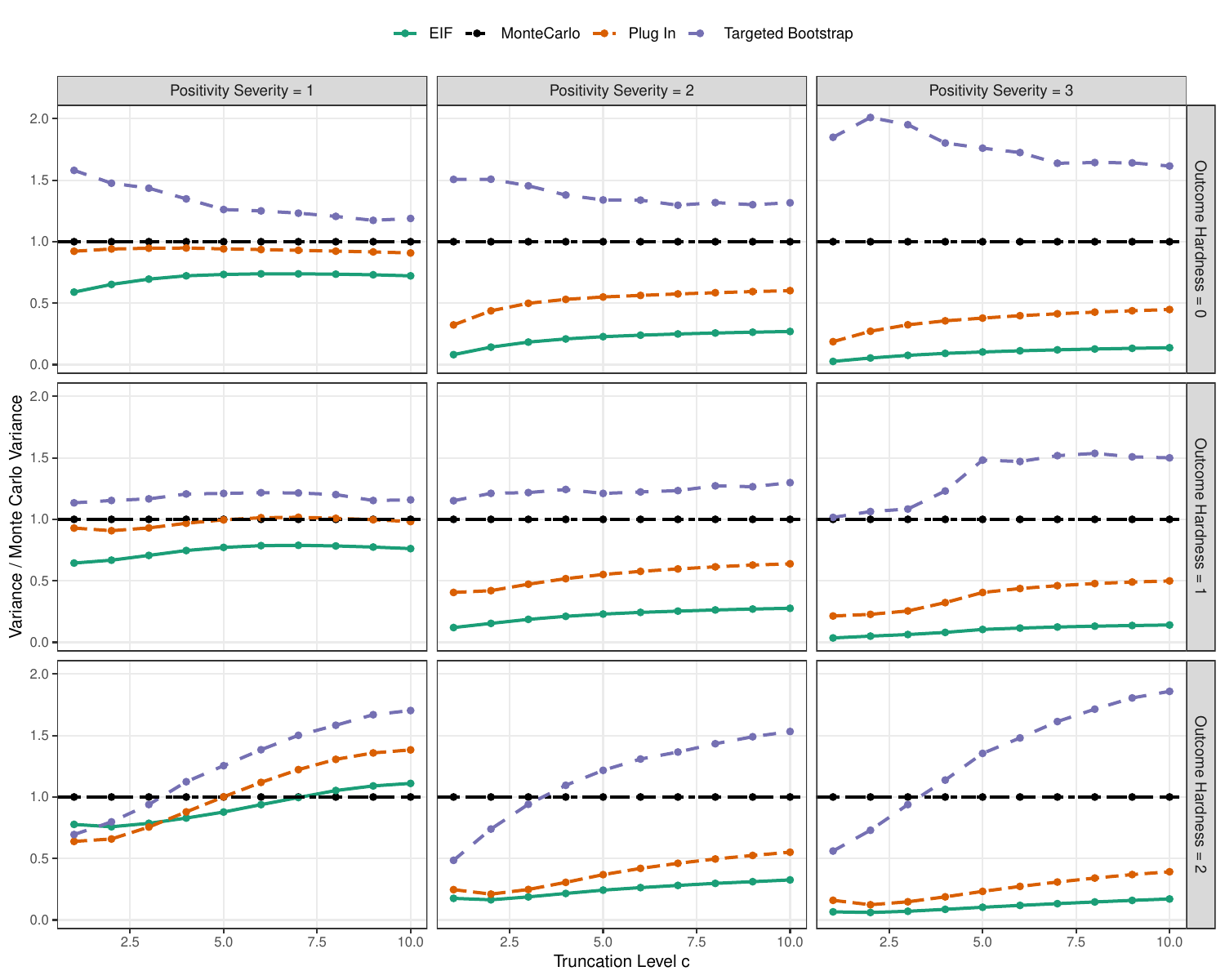}
    \caption{
    Relative variance estimates normalized by the Monte Carlo variance ($=1$) across truncation levels $c$.
    Rows correspond to increasing outcome hardness, and columns correspond to increasing positivity severity.
    The EIF-based, plug-in, and targeted bootstrap variance estimators are shown relative to the Monte Carlo benchmark (dashed line).
    }
    \label{fig:relative-var-gH}
\end{figure}

\Cref{fig:relative-var-gH} reports the relative variance estimates for the clever-covariate–scaled targeting estimator (gH) when the sample size is \(n=1000\). Each curve shows the ratio between the estimated variance and the Monte Carlo variance across truncation levels \(c\), with the Monte Carlo variance normalized to one. The panels vary by the severity of outcome model misspecification and the strength of the positivity violation parameter \(\kappa_{\mathrm{pos}}\). Numerical values of the variance estimates are reported in the appendix.

To illustrate the behavior of the variance estimators, consider three representative truncation levels \(c=3,6,10\) in the setting with high outcome misspecification and severe positivity violations (\(\kappa_{\mathrm{pos}}=3\)). The Monte Carlo variances (serving as the gold standard) are \(0.367\), \(0.106\), and \(0.046\), respectively. The EIF-based estimator substantially underestimates these values, reporting \(0.026\), \(0.013\), and \(0.008\) at \(c=3,6,10\). The plug-in estimator improves over EIF by partially correcting this downward bias, producing estimates \(0.054\), \(0.029\), and \(0.018\), which are consistently closer to the Monte Carlo benchmark. The targeted bootstrap (TB) estimator provides an additional stable alternative, yielding \(0.345\), \(0.157\), and \(0.086\) at \(c=3,6,10\), which track the magnitude of the Monte Carlo variance more closely across truncation levels.

Overall, the EIF estimator exhibits the strongest downward bias, while the plug-in estimator improves substantially over EIF by incorporating the conditional variance component. The targeted bootstrap estimator further stabilizes variance estimation along the truncation path and generally produces estimates that are closer to the Monte Carlo benchmark. In practice, TB provides a conservative and safe choice for inference: although it can sometimes overestimate the variance, especially when positivity violations are mild or outcome model misspecification is small, such conservativeness is preferable to the severe underestimation observed with the EIF estimator, as it helps maintain valid statistical inference. The appendix additionally reports the corresponding variance comparisons for the loss-weighted targeting estimator (gWt). Although the discrepancies between variance estimators are less pronounced in that case, the same qualitative pattern remains: the EIF estimator tends to underestimate the variance the most, whereas the plug-in and targeted bootstrap estimators provide more stable estimates across truncation levels.

\section{Discussion}

This work compares two TMLE targeting strategies under practical positivity violations: loss-weighted targeting and clever-covariate–scaled targeting. Although both approaches ultimately aim to solve the same efficient influence function estimating equation, they restrict the fluctuation step to different paths. Loss-weighted targeting applies inverse probability factors directly to the loss function, producing globally regularized updates of the outcome regression that do not depend on the magnitude of inverse probability weights. In contrast, clever-covariate–scaled targeting incorporates these factors through the clever covariate, allowing stronger local corrections in strata with large inverse probability weights. Empirically, this difference leads to a systematic bias–variance contrast: loss-weighted targeting produces smoother updates but accumulates substantial bias when overlap is poor, whereas clever-covariate–scaled targeting adapts corrections to high-weight regions and remains comparatively stable under outcome misspecification and practical positivity violations.

Our simulations further demonstrate that the truncation level plays a critical role in stabilizing clever-covariate–scaled targeting. Insufficient truncation produces extremely large inverse probability weights, leading to unstable estimates and inflated variance, while overly aggressive truncation introduces bias by distorting the effective target population. Importantly, the optimal truncation level depends on the sample size: more aggressive truncation tends to perform best in smaller samples, whereas less aggressive truncation becomes preferable as the sample size increases. At the same time, the recommended fixed choices $c=5$ and $c=6$ remain robust practical defaults across a wide range of settings. The pattern also motivates data-adaptive truncation as an alternative beyond relying on a single fixed rule. The proposed Lepski-type procedure with a brake mechanism advances along the truncation path only when successive estimator changes exceed the associated statistical uncertainty, while the brake prevents movement toward less truncation when the variance estimates become unreliable. In our simulations, the procedure often avoided the most unstable low-truncation regimes and provided competitive performance relative to fixed truncation rules, although it was not uniformly superior to recommended fixed choices such as $c=5$ or $c=6$.

Finally, we investigated variance estimation for inference under practical positivity violations. The efficient influence function–based variance estimator often substantially underestimates variability when inverse probability weights are extreme. A plug-in estimator partially corrects this behavior by incorporating conditional outcome variance, while the targeted bootstrap provides a stable alternative that more closely tracks the Monte Carlo variability across truncation levels. Although the targeted bootstrap may occasionally overestimate variance in milder settings, its conservative behavior helps maintain valid inference when positivity violations are severe.

\printbibliography

\clearpage 
\newpage
\appendix
\setcounter{figure}{0}
\renewcommand{\thefigure}{A.\arabic{figure}}

\setcounter{table}{0}
\renewcommand{\thetable}{A.\arabic{table}}
\section*{Appendix}
\label{sec:app}

\section*{Auxiliary theoretical analysis}

\begin{proof}[Proof. \textbf{Plug-in variance estimation}]
Recall the efficient influence function for the ATE:
\begin{align*}
D(O)
&=
\sum_{a\in\{0,1\}}
\frac{I(A=a)}{g(a|W)}(Y-\bar Q(a,W))
+
\bar Q(1,W)-\bar Q(0,W)-\Psi.
\end{align*}

We decompose its variance:
\begin{align*}
\mathrm{Var}(D(O))
&=
\sum_{a\in\{0,1\}}
\mathrm{Var}\!\left(
\frac{I(A=a)}{g(a|W)}(Y-\bar Q(a,W))
\right)
+
\mathrm{Var}\!\left(
\bar Q(1,W)-\bar Q(0,W)-\Psi
\right),
\end{align*}
since the cross-covariance terms vanish. Indeed,
\begin{align*}
E\!\left[
I(A=a)(Y-\bar Q(a,W)) \mid W
\right]
&=
g(a|W)
E\!\left[
Y-\bar Q(a,W) \mid A=a,W
\right] \\
&=
0.
\end{align*}

For continuous outcomes,
\begin{align*}
E\!\left[
\left(
\frac{I(A=a)}{g(a|W)}
\right)^2
(Y-\bar Q(a,W))^2
\right]
&=
E\!\left[
\frac{I(A=a)}{g(a|W)^2}
(Y-\bar Q(a,W))^2
\right] \\
&=
E_W\!\left[
\frac{\mathrm{Var}(Y \mid A=a,W)}{g(a|W)}
\right].
\end{align*}

Let $R_i = Y_i - \bar Q(A_i,W_i)$. Then
\begin{align*}
E[R_i^2 \mid A=a,W]
&=
\mathrm{Var}(Y \mid A=a,W).
\end{align*}

Estimating this conditional variance by regressing $R_i^2$ on $W$ within each treatment arm yields $\hat\sigma^2(a,W)$. Substituting these estimates into the variance decomposition gives
\begin{align*}
\widehat{\mathrm{Var}}_{\text{plug-in}}(\hat\psi)
&=
\frac{1}{n}
\sum_{i=1}^{n}
\left[
\frac{\hat\sigma^2(1,W_i)}{\hat g(1|W_i)}
+
\frac{\hat\sigma^2(0,W_i)}{\hat g(0|W_i)}
\right] \\
&\quad +
\frac{1}{n}
\sum_{i=1}^n
\left[
\hat Q(1,W_i)-\hat Q(0,W_i)-\hat\psi
\right]^2,
\end{align*}
which establishes the stated plug-in variance representation.
\end{proof}

\section*{Detailed and additional simulation results}

\begin{figure}[t!]
  \centering
  \includegraphics[width=0.95\textwidth]{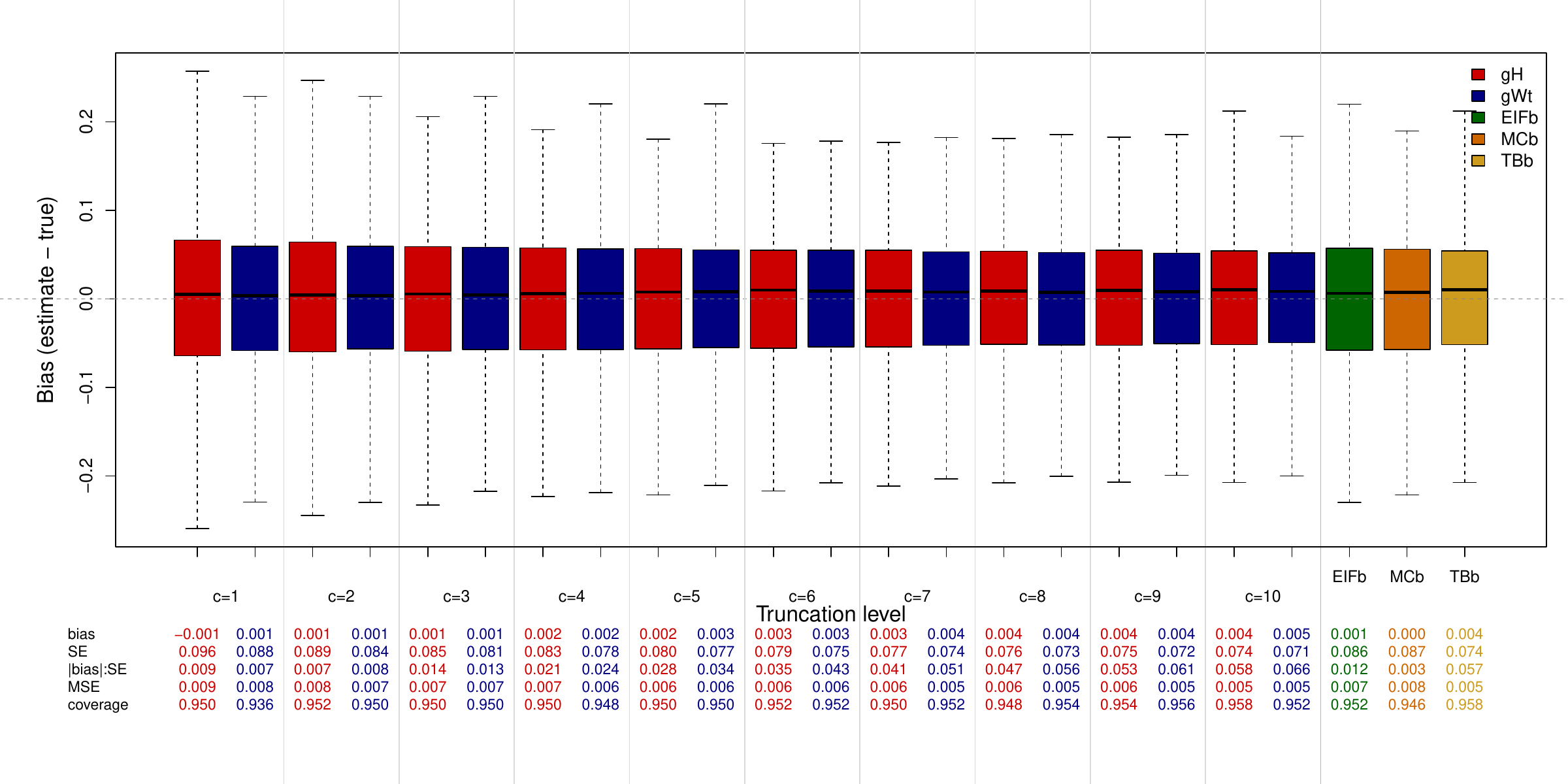}
  \caption{EIFb/MCb/TBb adaptive truncation (n=500): nearly correct outcome, mild positivity.}
  \label{fig:n500_mu0pos1}
\end{figure}

\begin{figure}[t!]
  \centering
  \includegraphics[width=0.95\textwidth]{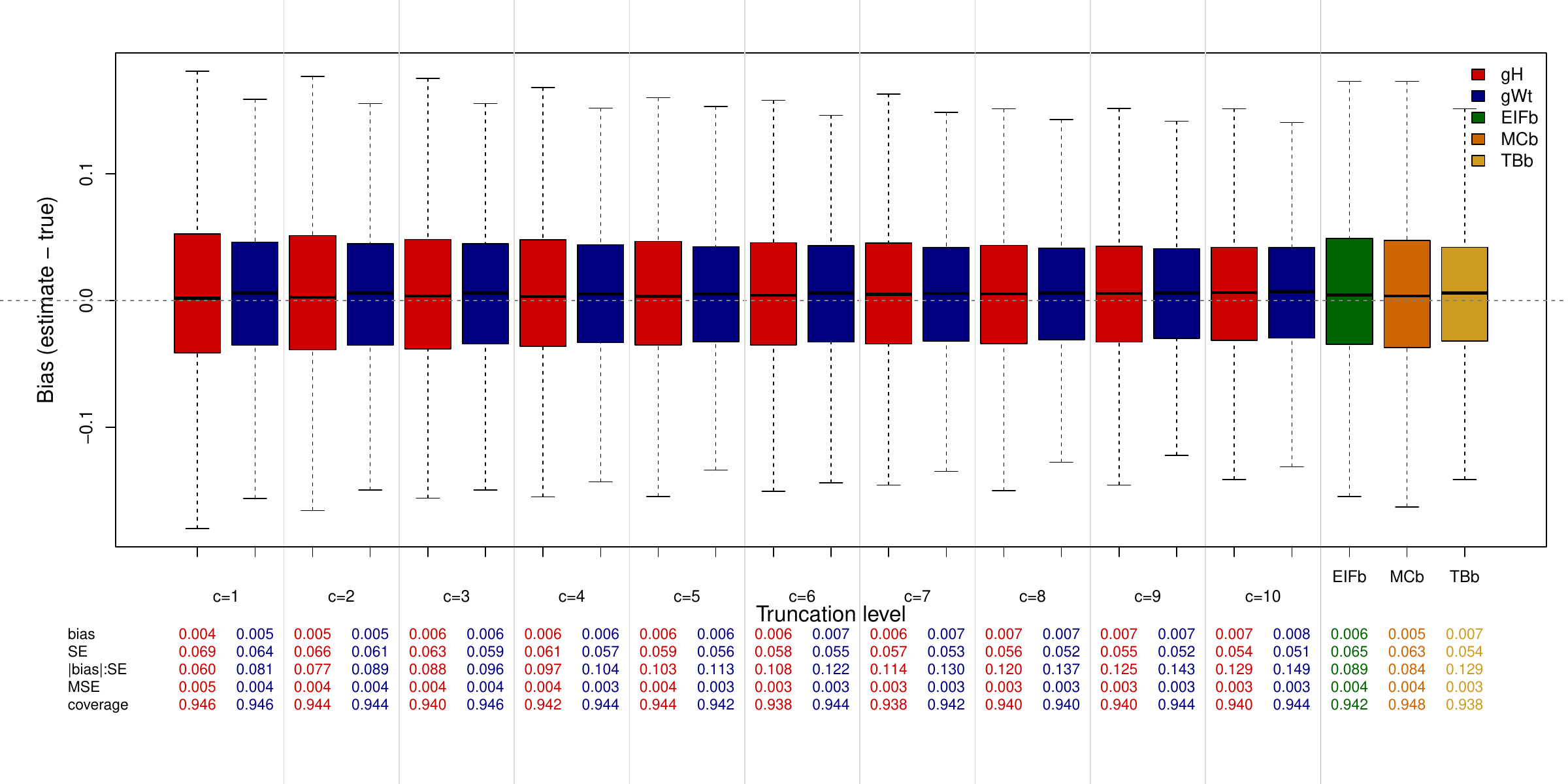}
  \caption{EIFb/MCb/TBb adaptive truncation (n=1000): nearly correct outcome, mild positivity.}
  \label{fig:n1000_mu0pos1}
\end{figure}

\begin{figure}[t!]
  \centering
  \includegraphics[width=0.95\textwidth]{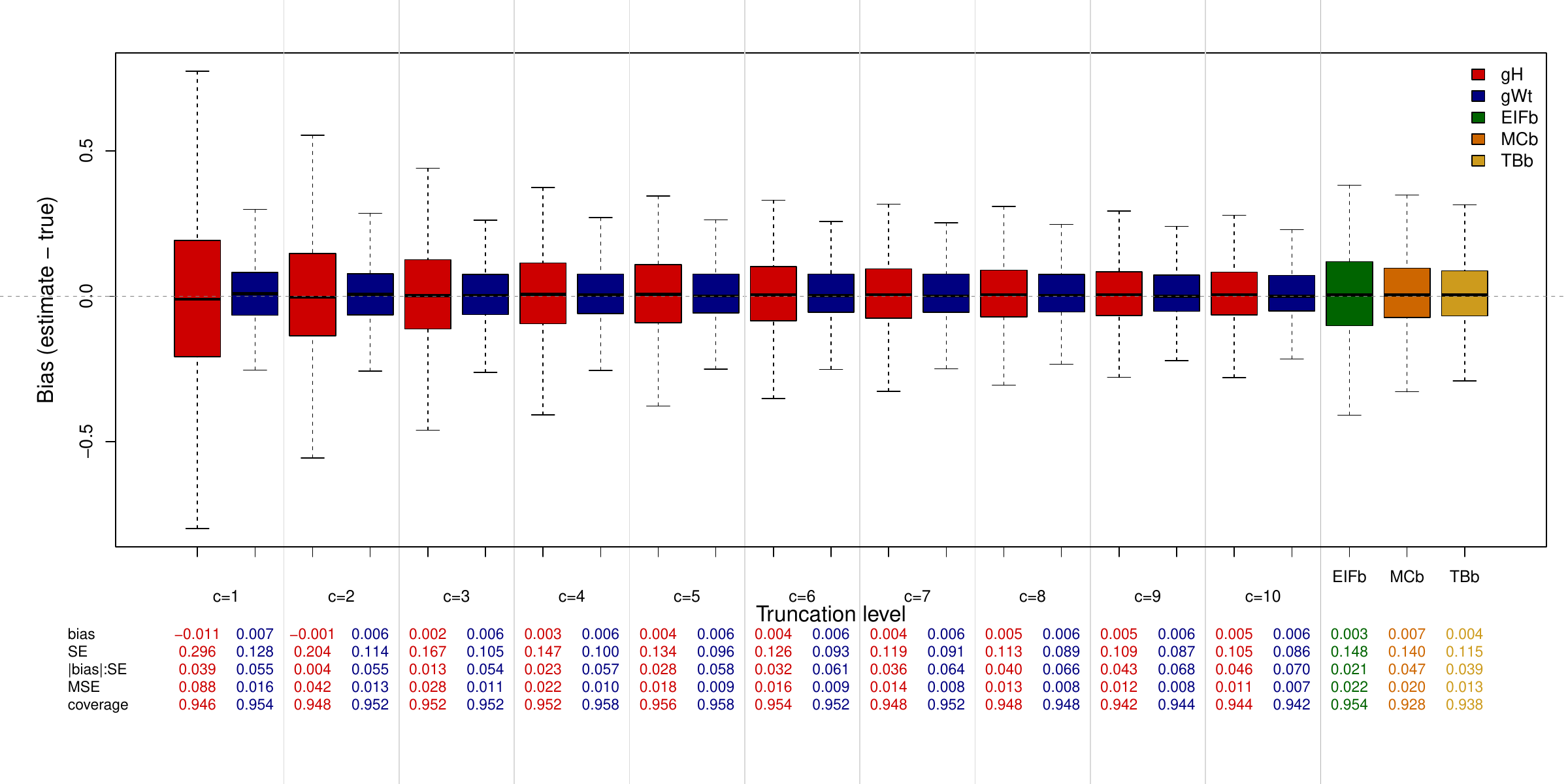}
  \caption{EIFb/MCb/TBb adaptive truncation (n=500): nearly correct outcome, moderate positivity.}
  \label{fig:n500_mu0pos2}
\end{figure}

\begin{figure}[t!]
  \centering
  \includegraphics[width=0.95\textwidth]{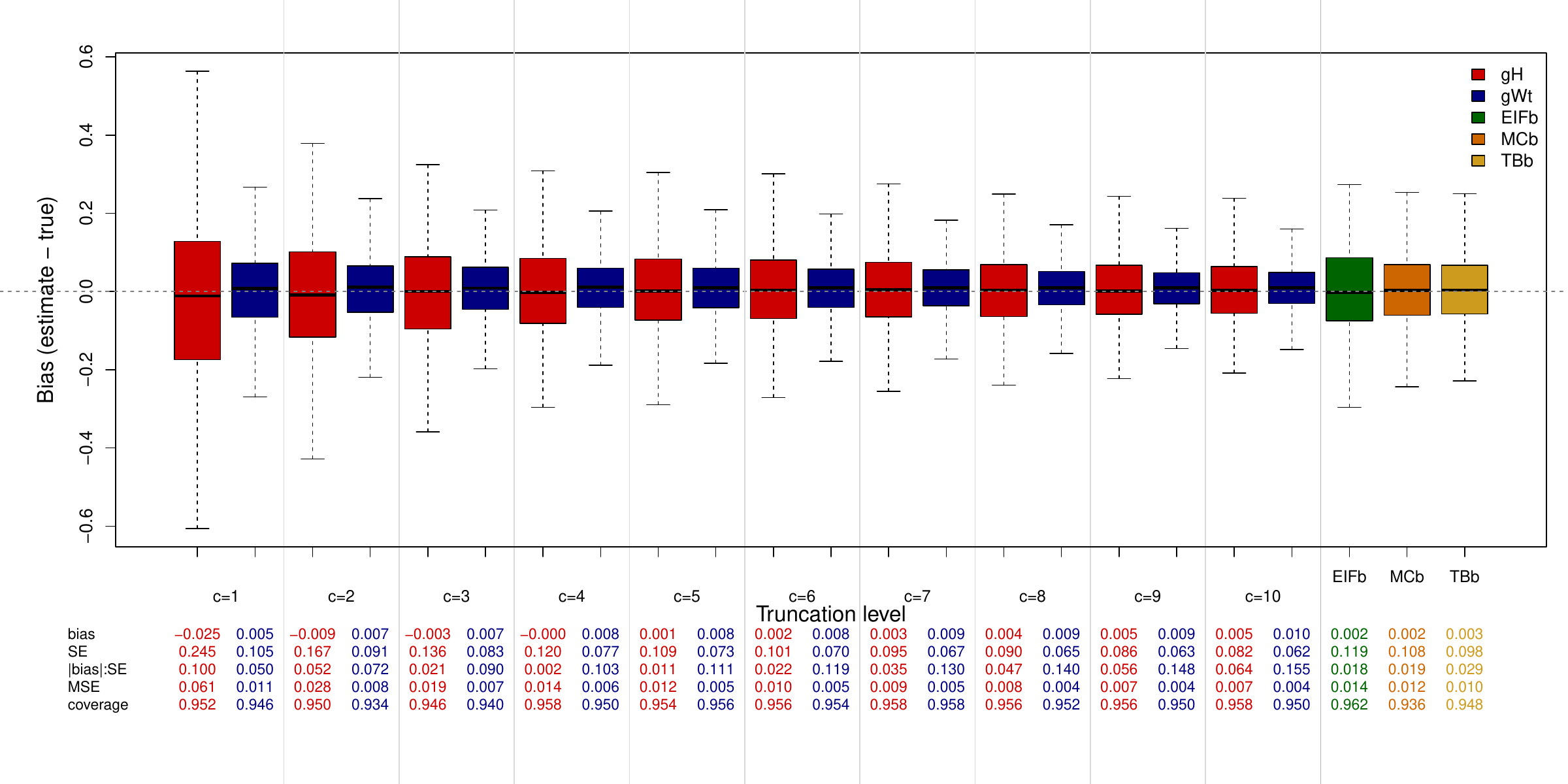}
  \caption{EIFb/MCb/TBb adaptive truncation (n=1000): nearly correct outcome, moderate positivity.}
  \label{fig:n1000_mu0pos2}
\end{figure}

\begin{figure}[t!]
  \centering
  \includegraphics[width=0.95\textwidth]{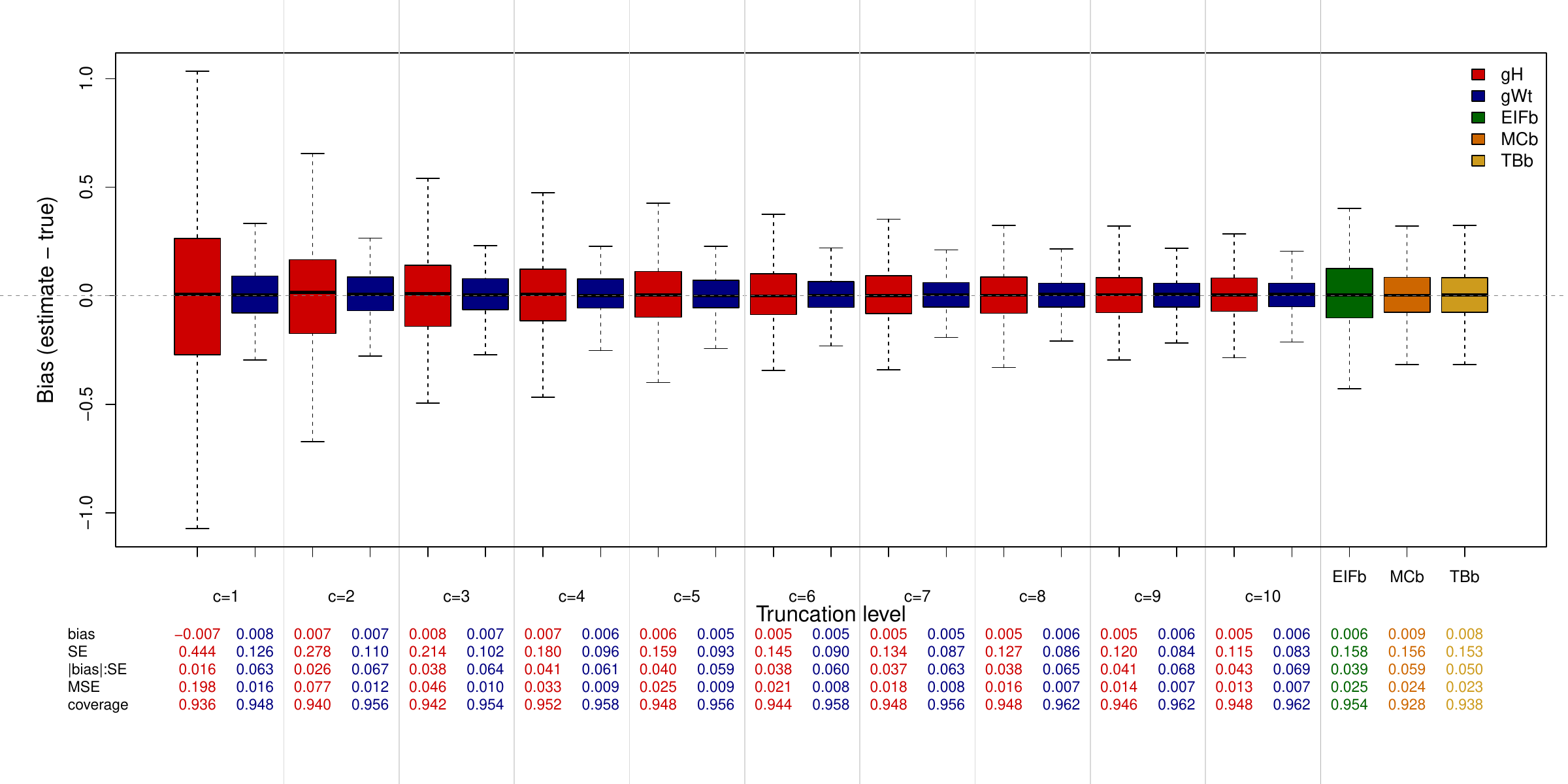}
  \caption{EIFb/MCb/TBb adaptive truncation (n=500): nearly correct outcome, severe positivity.}
  \label{fig:n500_mu0pos3}
\end{figure}

\begin{figure}[t!]
  \centering
  \includegraphics[width=0.95\textwidth]{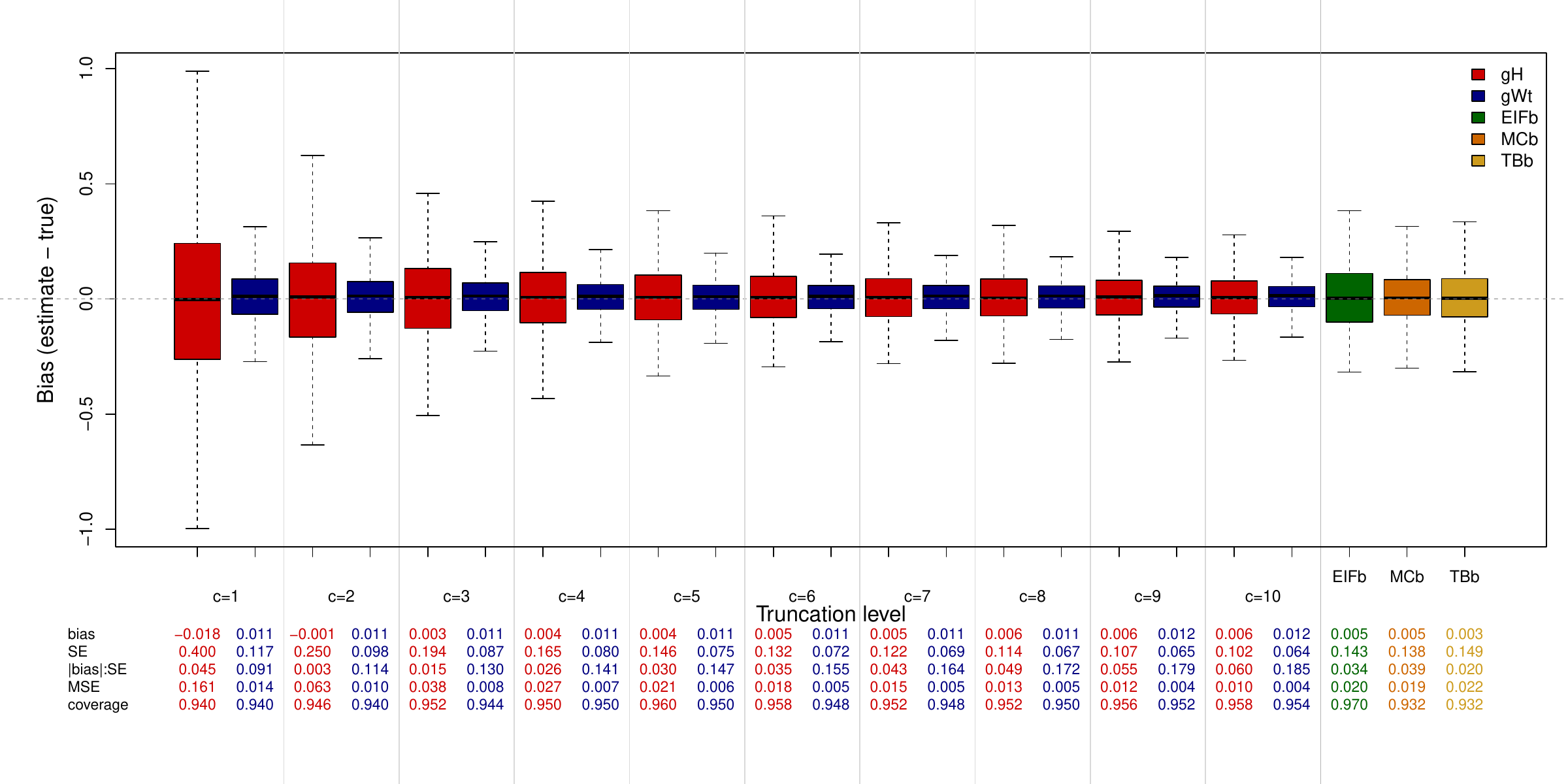}
  \caption{EIFb/MCb/TBb adaptive truncation (n=1000): nearly correct outcome, severe positivity.}
  \label{fig:n1000_mu0pos3}
\end{figure}

\begin{figure}[t!]
  \centering
  \includegraphics[width=0.95\textwidth]{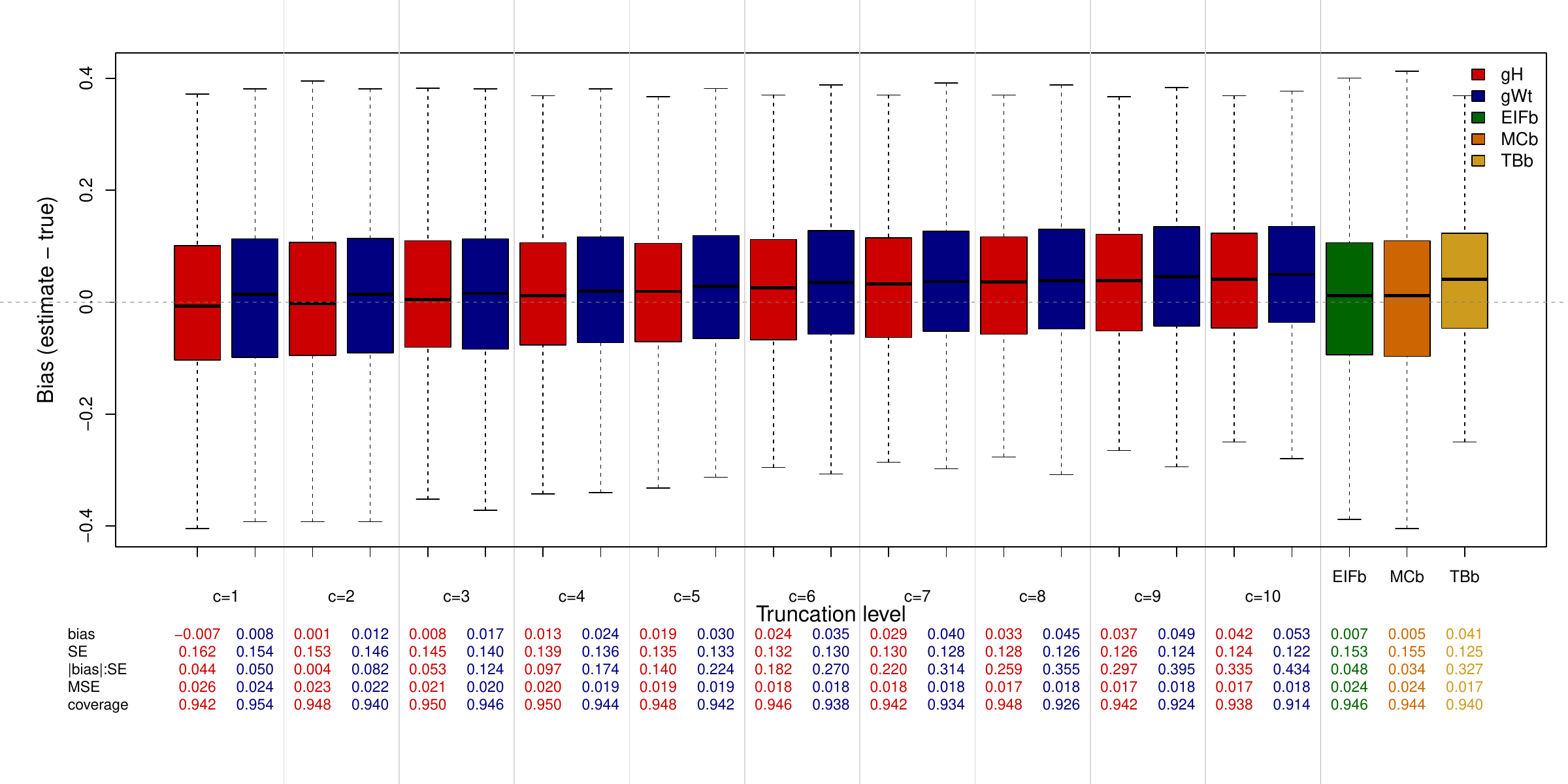}
  \caption{EIFb/MCb/TBb adaptive truncation (n=500): moderate misspecification, mild positivity.}
  \label{fig:n500_mu1pos1}
\end{figure}

\begin{figure}[t!]
  \centering
  \includegraphics[width=0.95\textwidth]{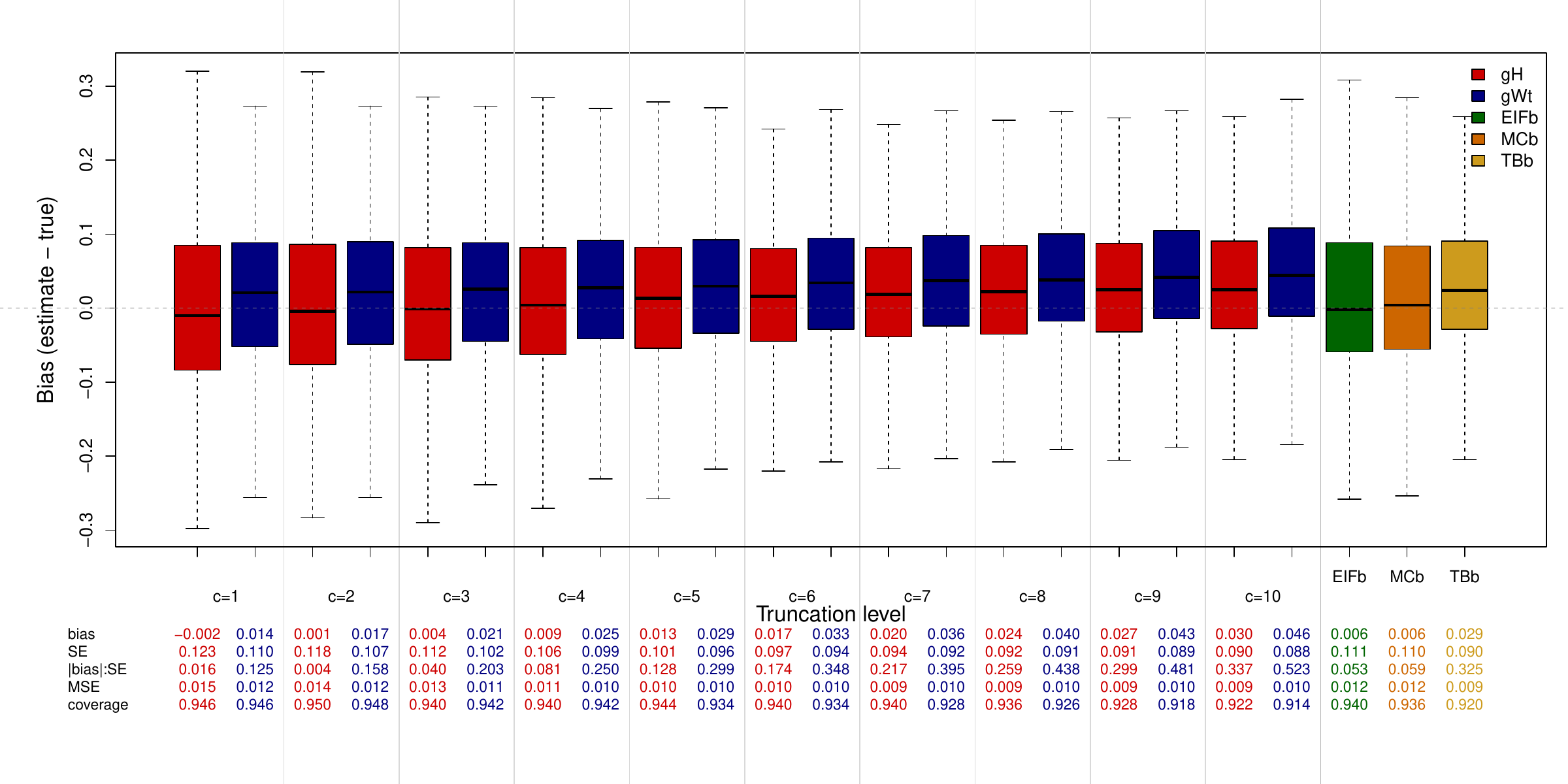}
  \caption{EIFb/MCb/TBb adaptive truncation (n=1000): moderate misspecification, mild positivity.}
  \label{fig:n1000_mu1pos1}
\end{figure}

\begin{figure}[t!]
  \centering
  \includegraphics[width=0.95\textwidth]{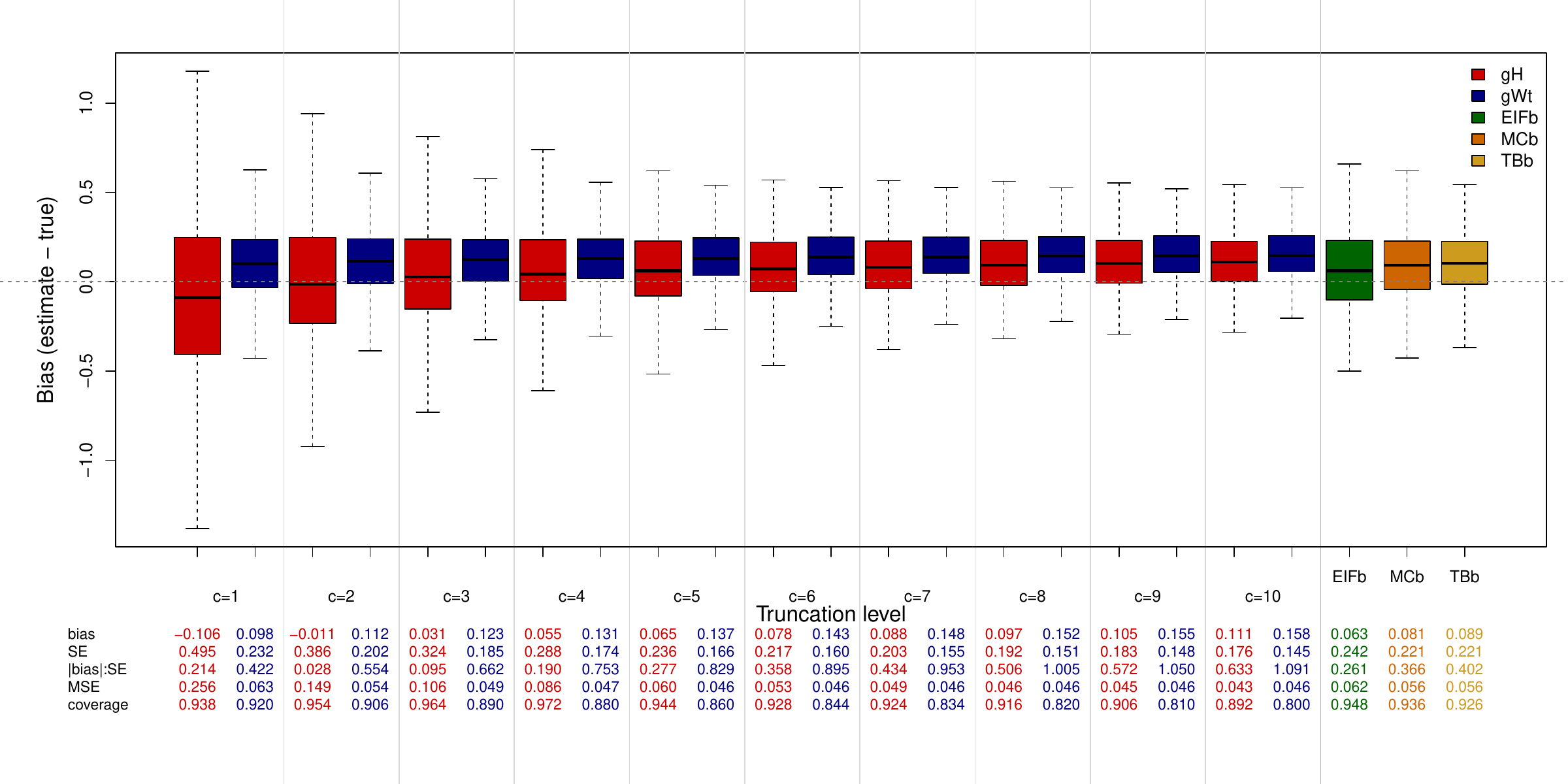}
  \caption{EIFb/MCb/TBb adaptive truncation (n=500): moderate misspecification, moderate positivity.}
  \label{fig:n500_mu1pos2}
\end{figure}

\begin{figure}[t!]
  \centering
  \includegraphics[width=0.95\textwidth]{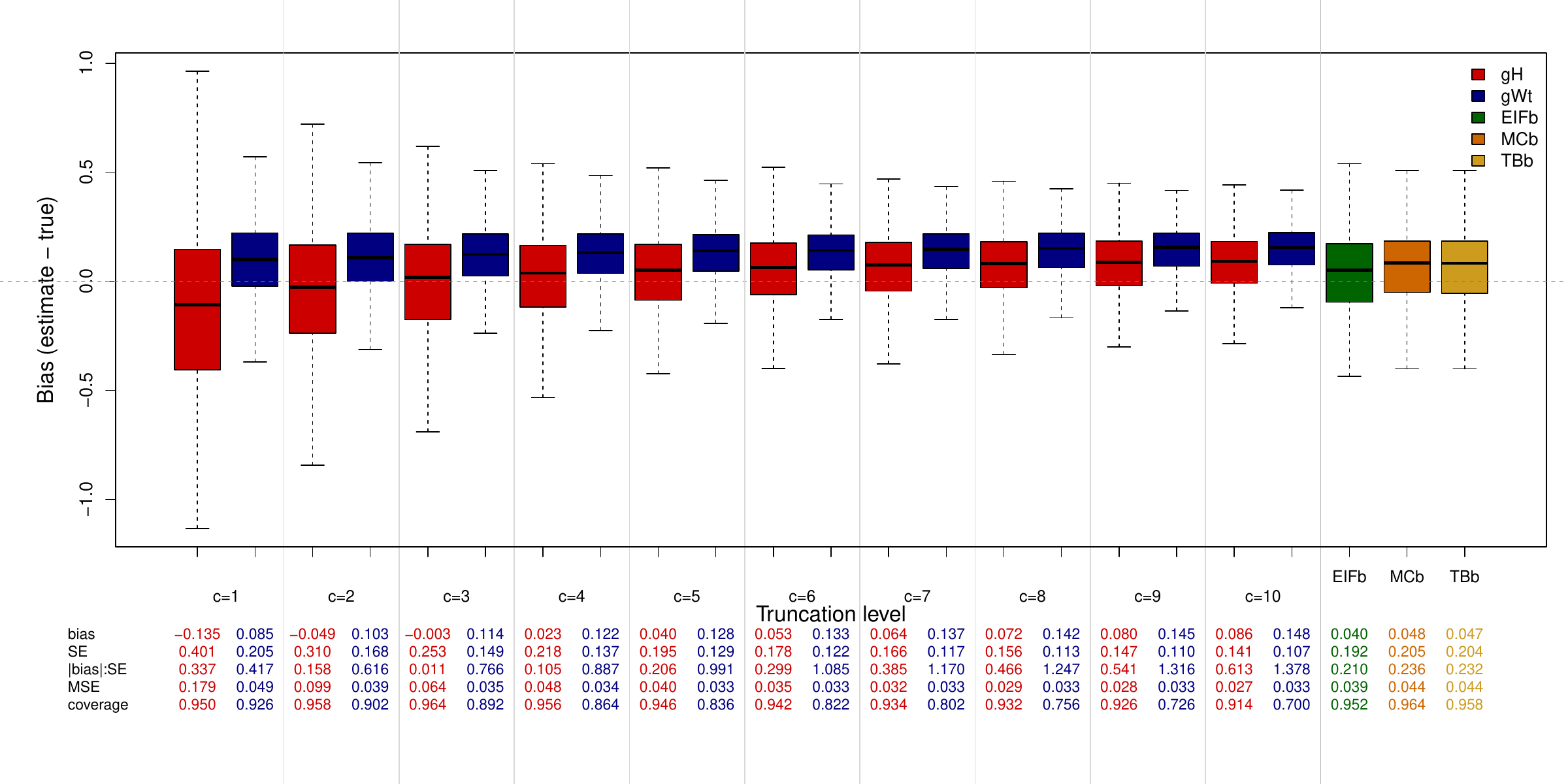}
  \caption{EIFb/MCb/TBb adaptive truncation (n=1000): moderate misspecification, moderate positivity.}
  \label{fig:n1000_mu1pos2}
\end{figure}

\begin{figure}[t!]
  \centering
  \includegraphics[width=0.95\textwidth]{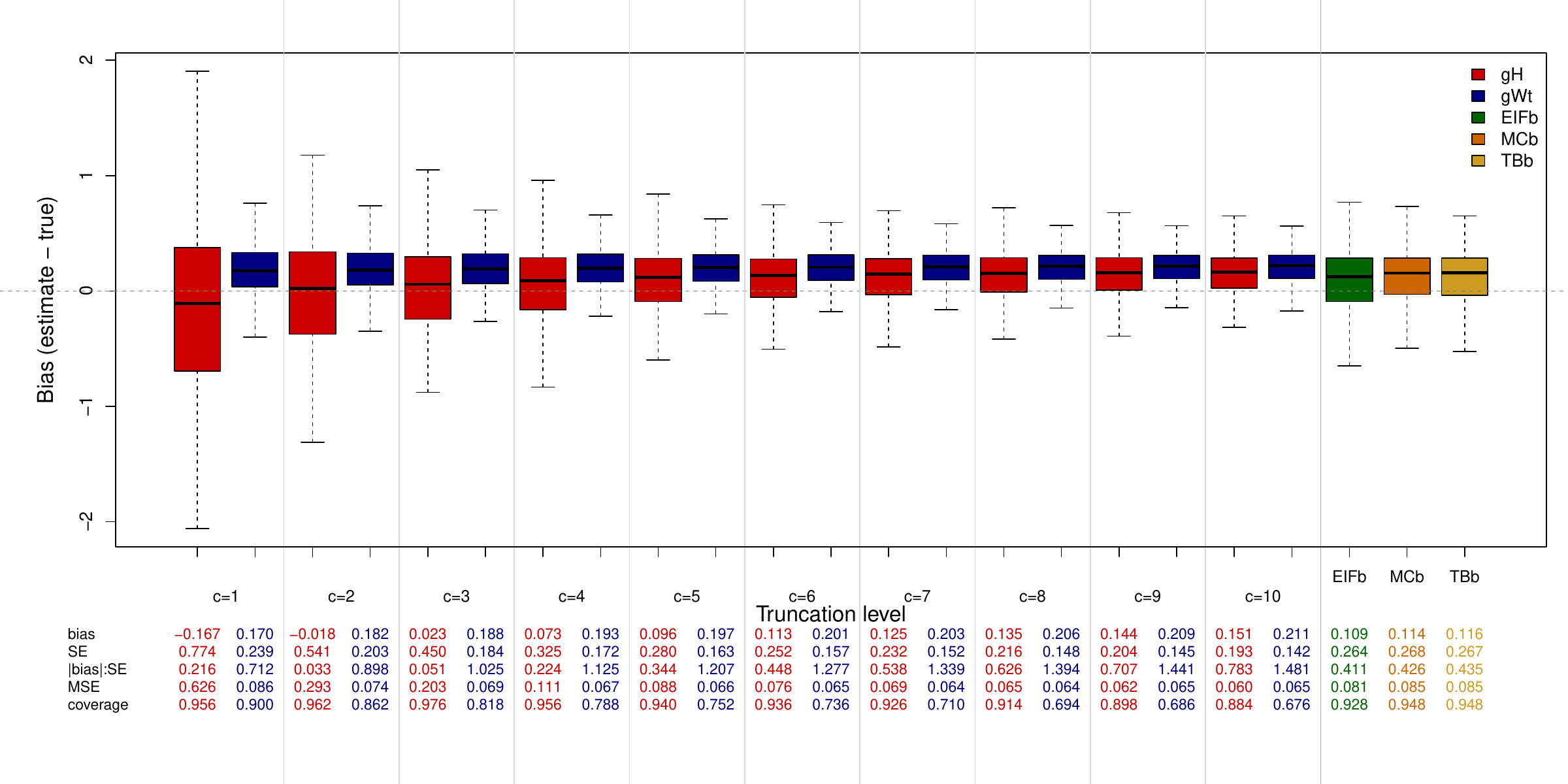}
  \caption{EIFb/MCb/TBb adaptive truncation (n=500): moderate misspecification, severe positivity.}
  \label{fig:n500_mu1pos3}
\end{figure}

\begin{figure}[t!]
  \centering
  \includegraphics[width=0.95\textwidth]{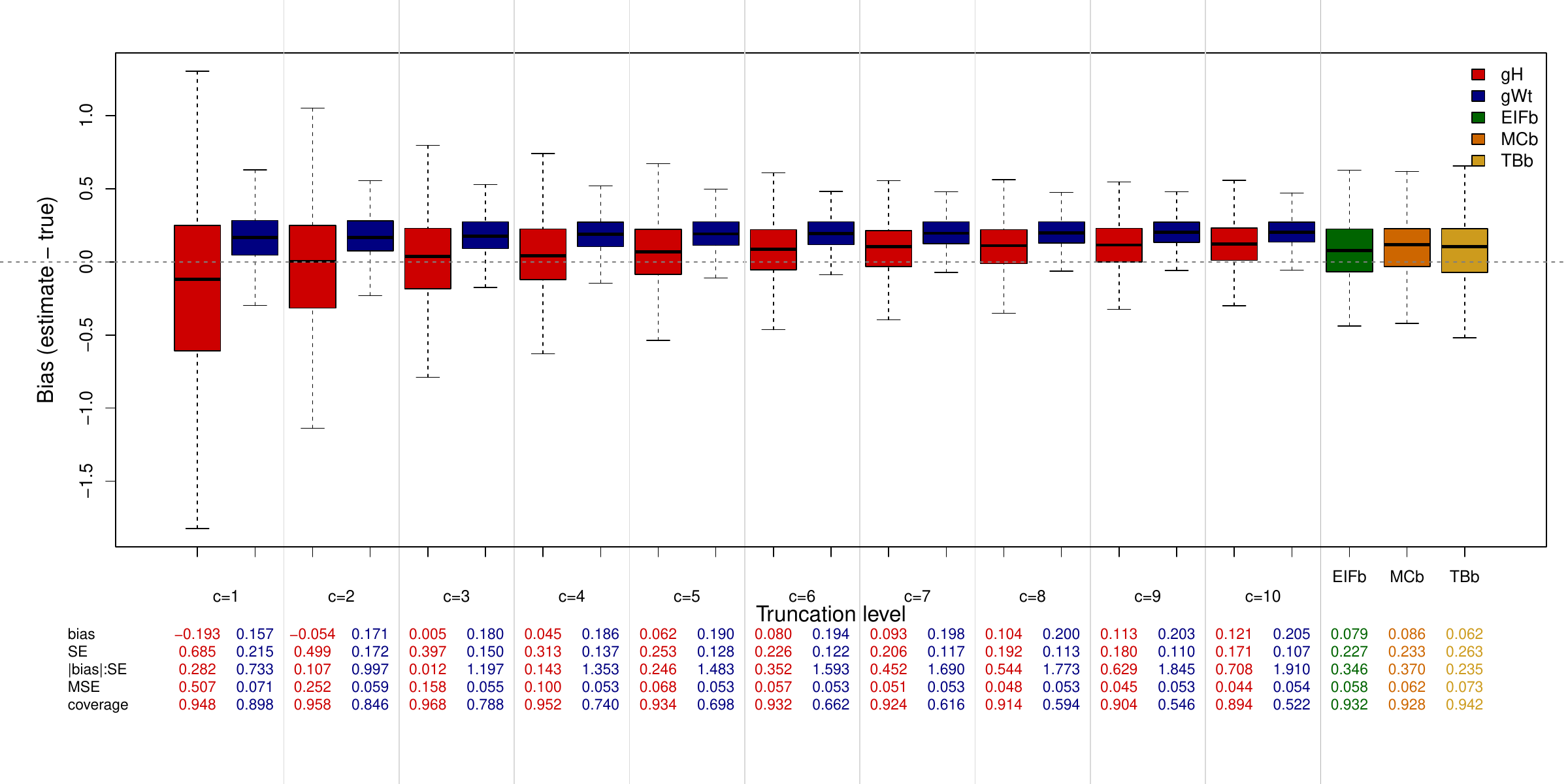}
  \caption{EIFb/MCb/TBb adaptive truncation (n=1000): moderate misspecification, severe positivity.}
  \label{fig:n1000_mu1pos3}
\end{figure}

\begin{figure}[t!]
  \centering
  \includegraphics[width=0.95\textwidth]{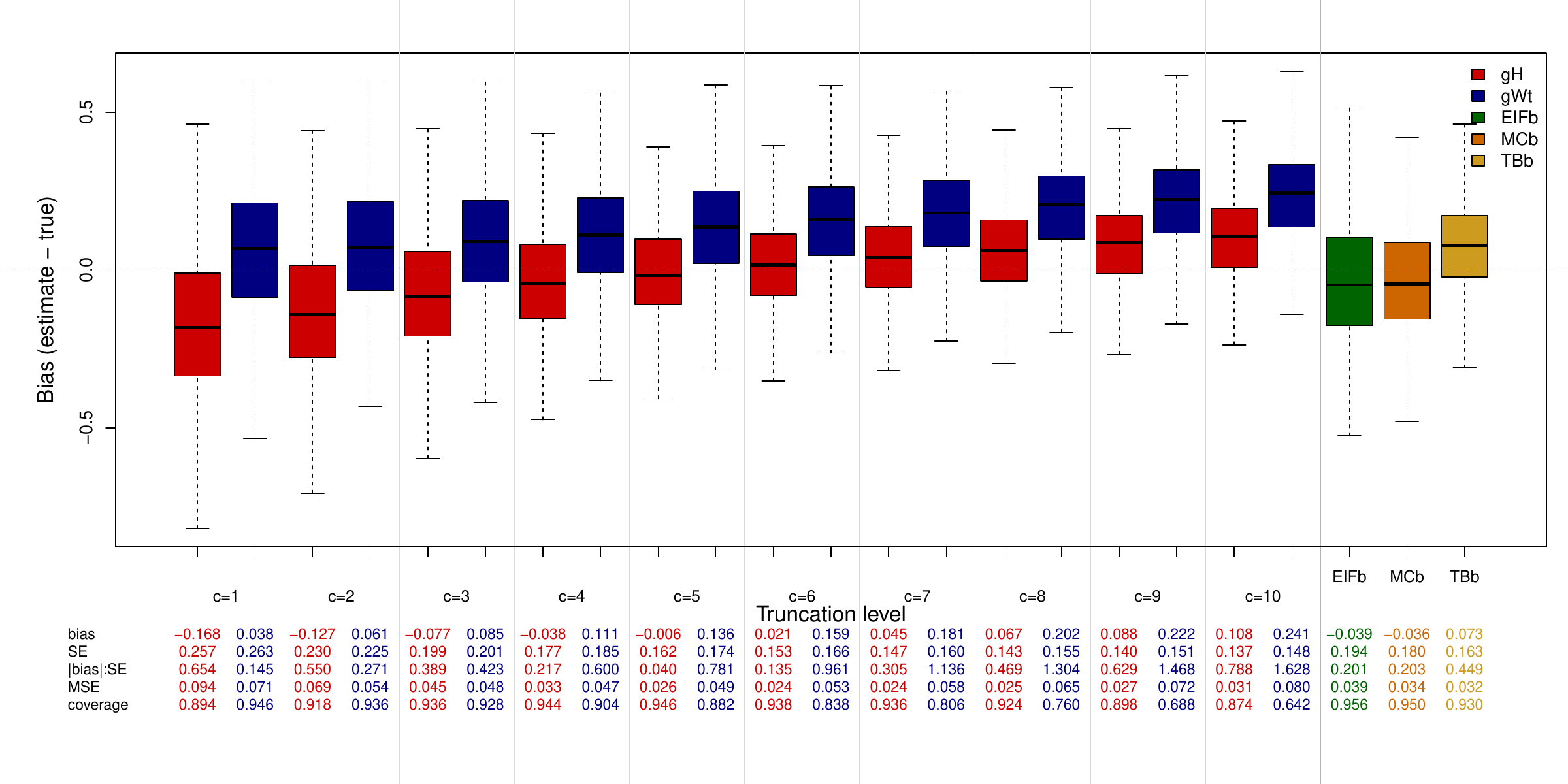}
  \caption{EIFb/MCb/TBb adaptive truncation (n=500): high misspecification, mild positivity.}
  \label{fig:n500_mu2pos1}
\end{figure}

\begin{figure}[t!]
  \centering
  \includegraphics[width=0.95\textwidth]{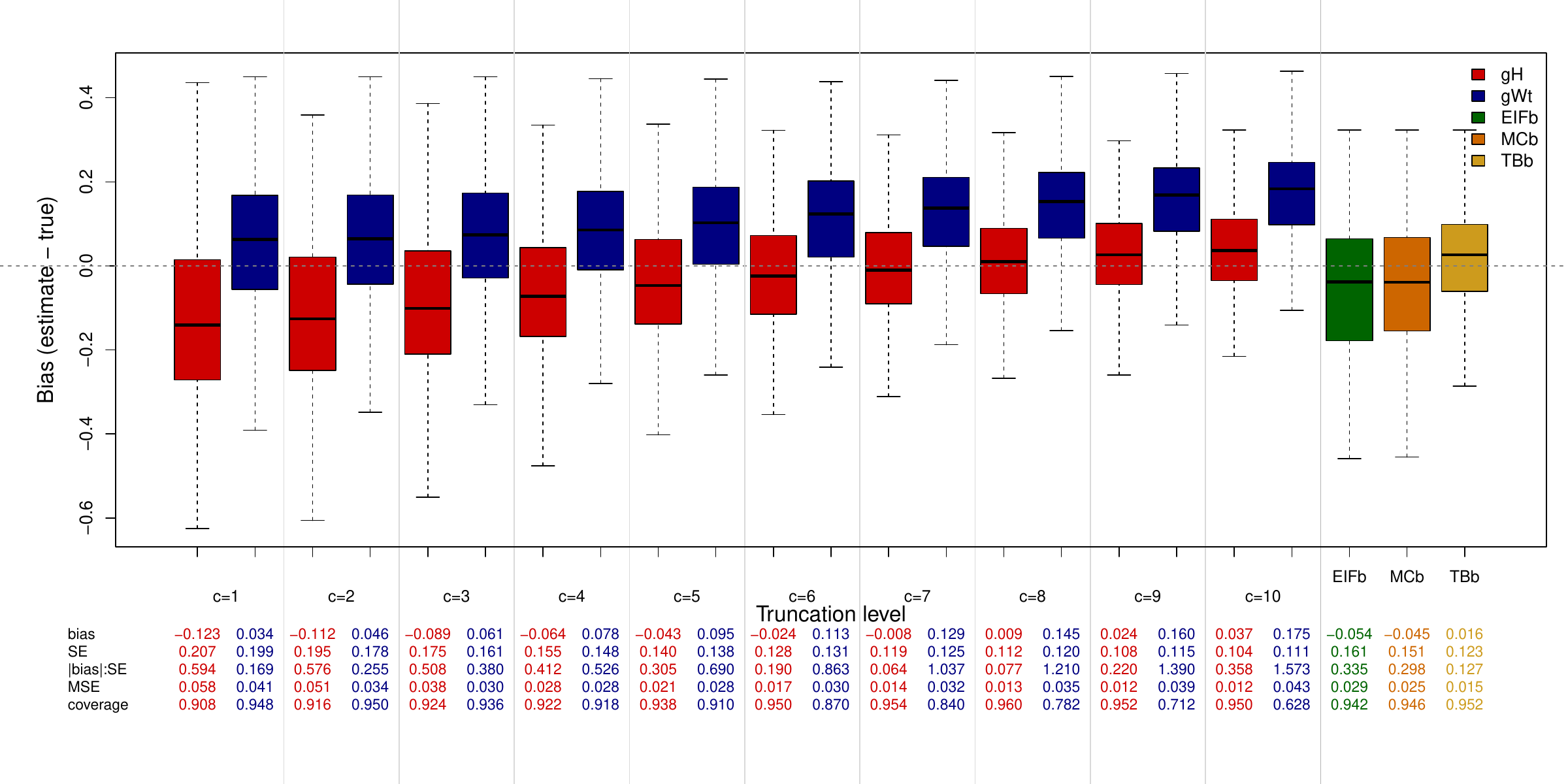}
  \caption{EIFb/MCb/TBb adaptive truncation (n=1000): high misspecification, mild positivity.}
  \label{fig:n1000_mu2pos1}
\end{figure}

\begin{figure}[t!]
  \centering
  \includegraphics[width=0.95\textwidth]{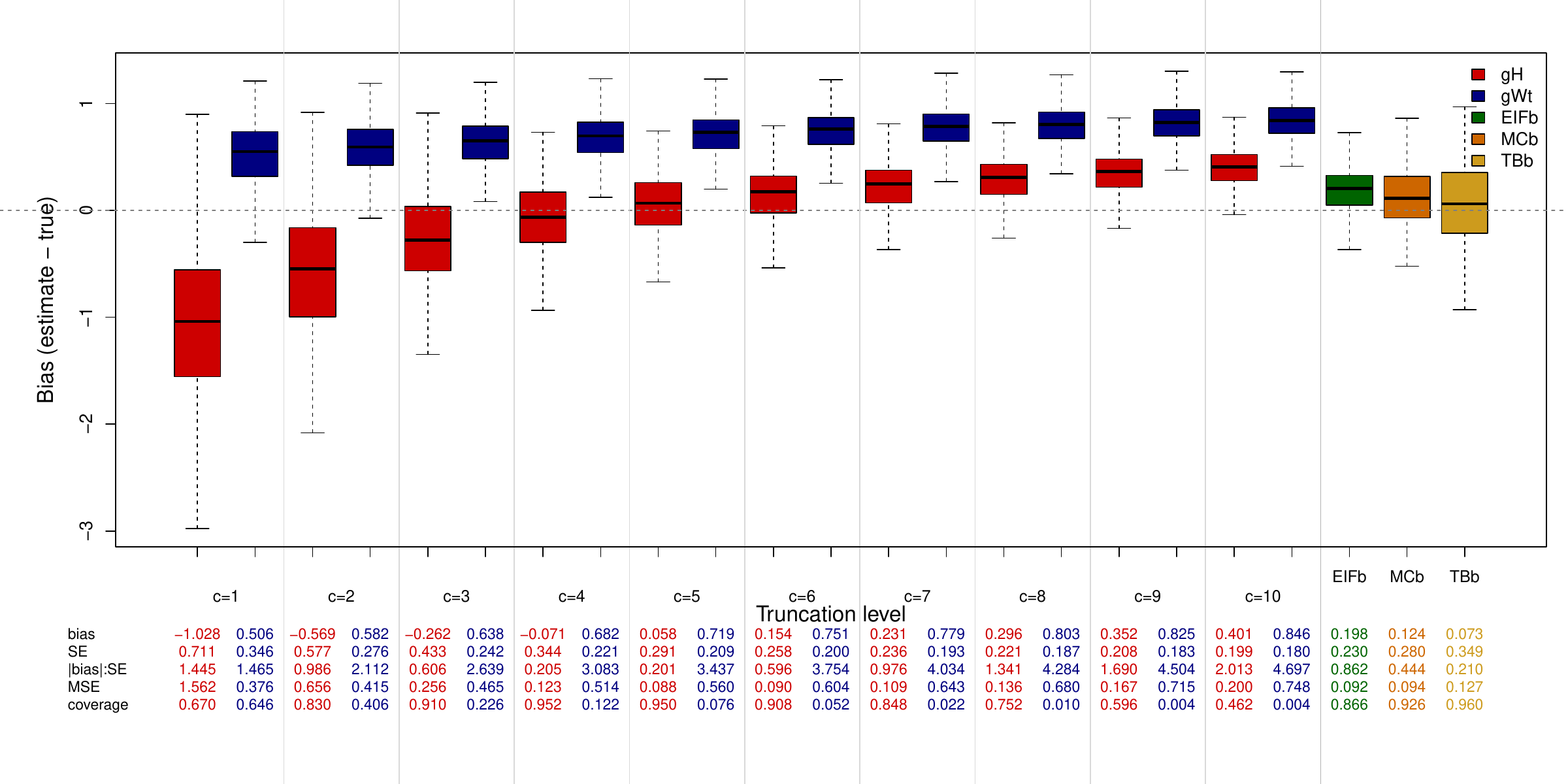}
  \caption{EIFb/MCb/TBb adaptive truncation (n=500): high misspecification, moderate positivity.}
  \label{fig:n500_mu2pos2}
\end{figure}

\begin{figure}[t!]
  \centering
  \includegraphics[width=0.95\textwidth]{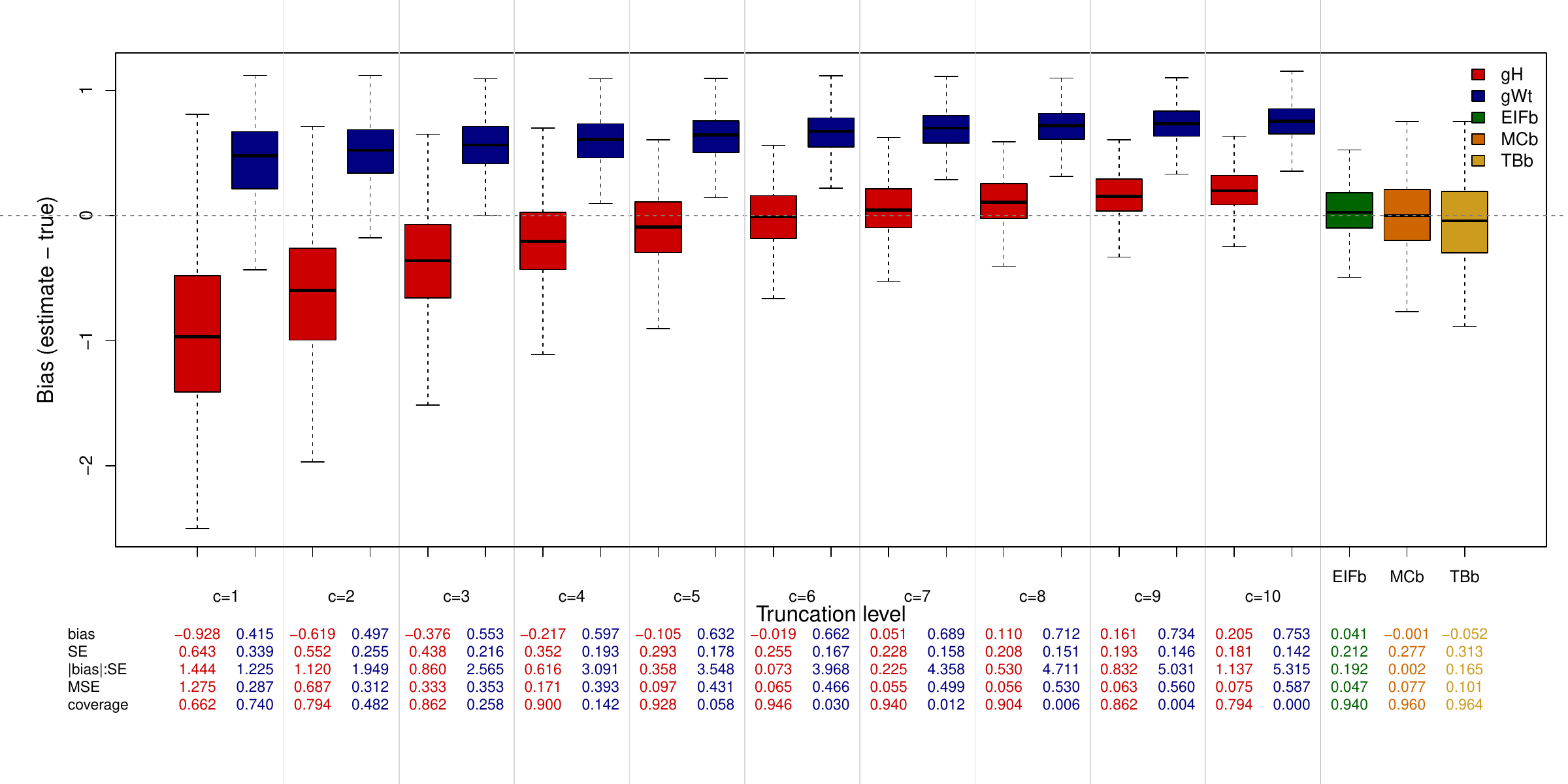}
  \caption{EIFb/MCb/TBb adaptive truncation (n=1000): high misspecification, moderate positivity.}
  \label{fig:n1000_mu2pos2}
\end{figure}

\begin{figure}[t!]
  \centering
  \includegraphics[width=0.95\textwidth]{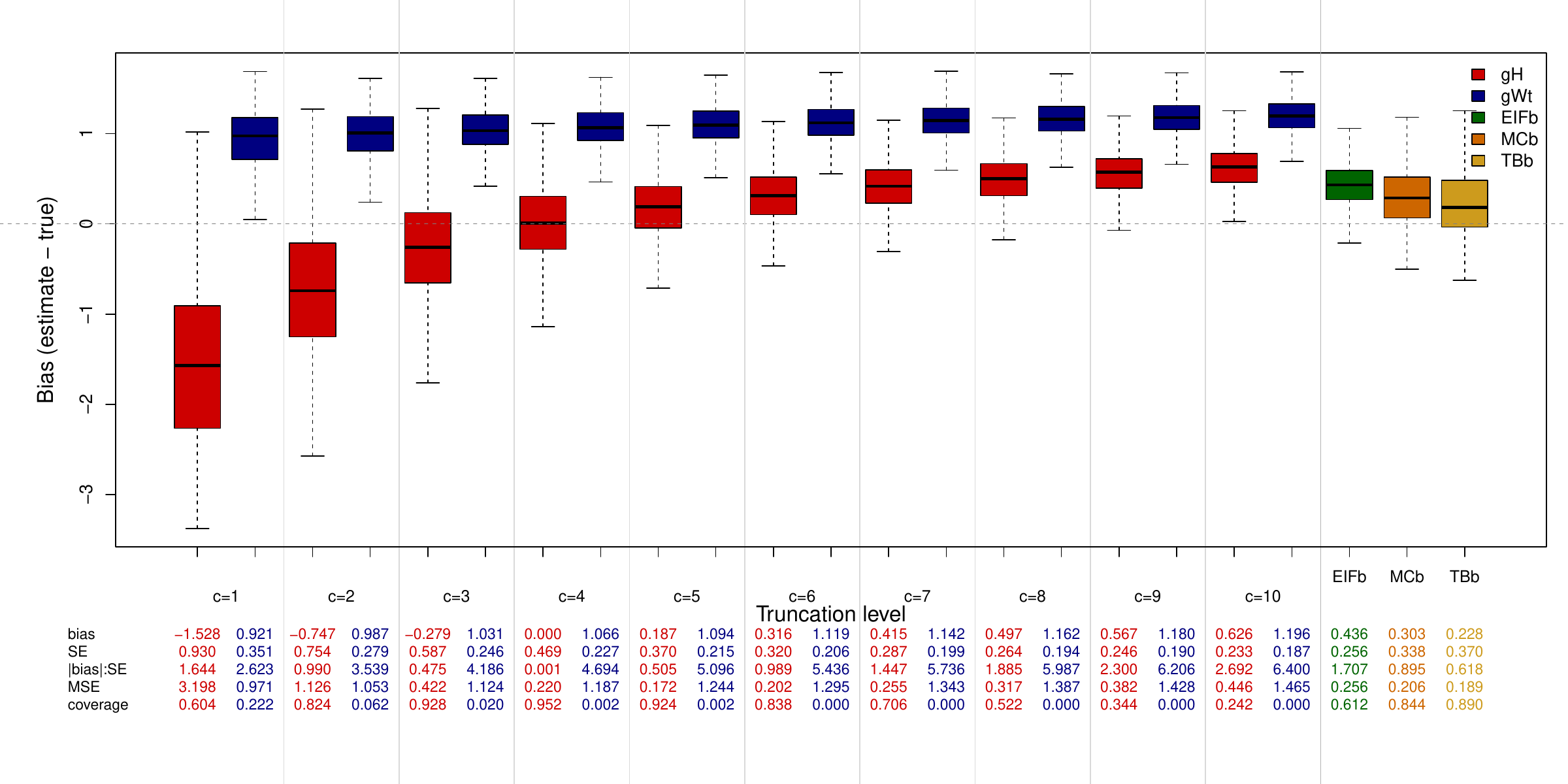}
  \caption{EIFb/MCb/TBb adaptive truncation (n=500): high misspecification, severe positivity.}
  \label{fig:n500_mu2pos3}
\end{figure}

\begin{figure}[t]
    \centering
    \includegraphics[width=0.9\textwidth]{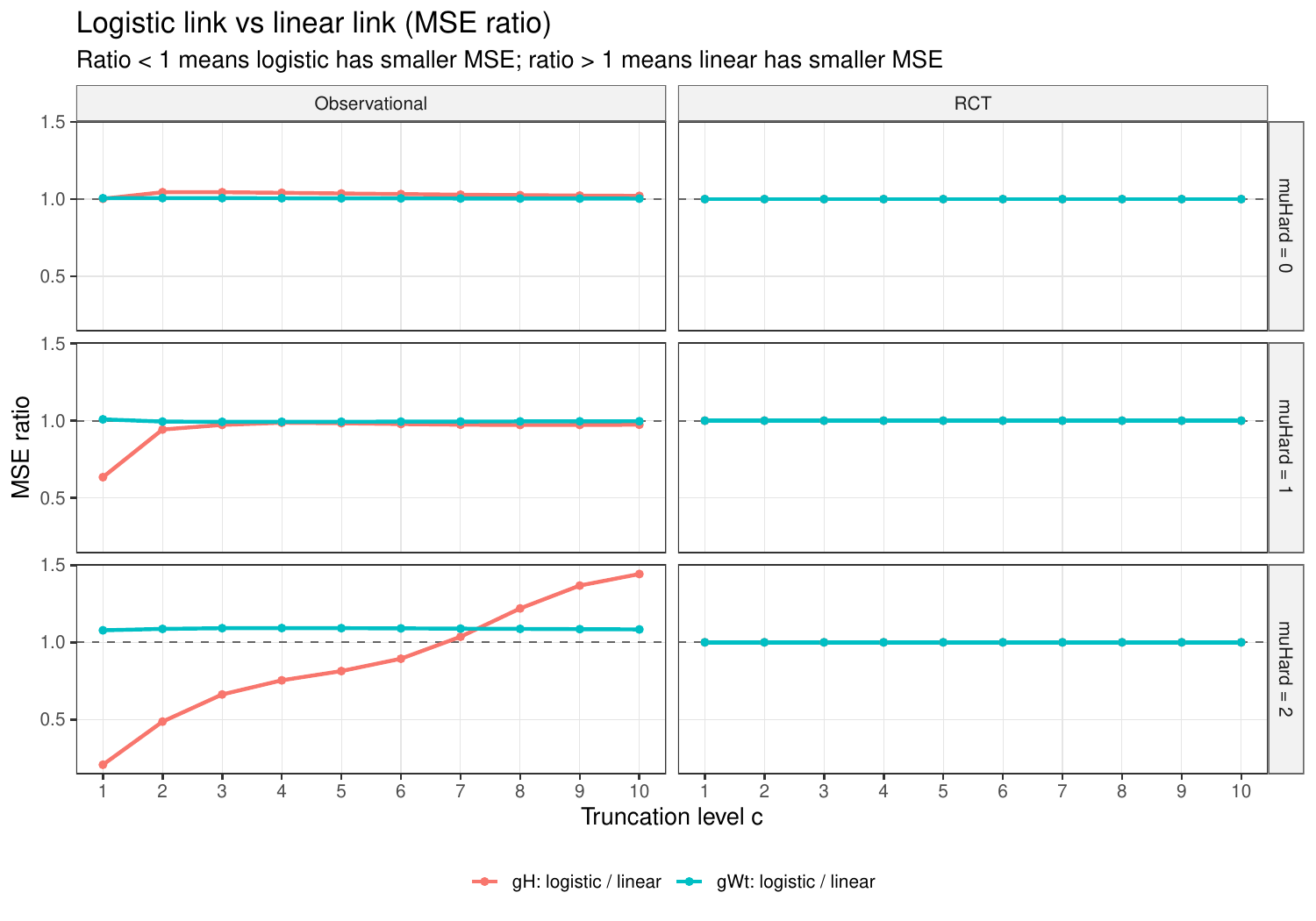}
    \caption{MSE ratio comparing the logistic link to the linear link across truncation levels $c=1,\dots,10$. Ratios below 1 indicate smaller MSE under the logistic link. Panels vary by outcome hardness ($\mu_{\text{hard}}$) and study type (observational vs.\ RCT).}
    \label{fig:mse-ratio-link}
\end{figure}

\begin{figure}[t]
    \centering
    \includegraphics[width=0.9\textwidth]{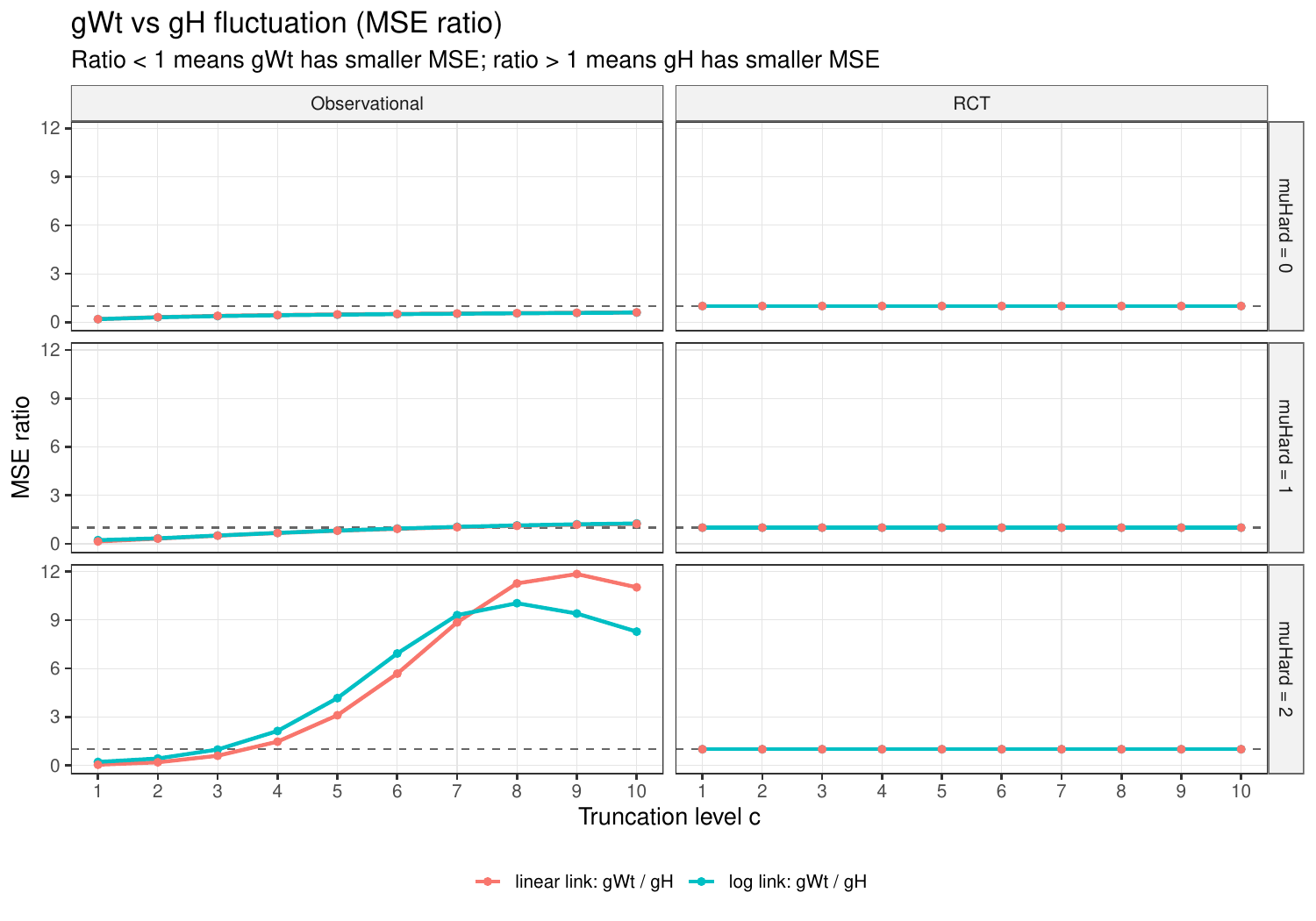}
    \caption{MSE ratio comparing $gWt$ to $gH$ fluctuation across truncation levels $c=1,\dots,10$, within each link function. Ratios below 1 indicate smaller MSE under $gWt$. Panels vary by outcome hardness ($\mu_{\text{hard}}$) and study type (observational vs.\ RCT).}
    \label{fig:mse-ratio-fluct}
\end{figure}

\begin{figure}[t]
    \centering
    \includegraphics[width=0.9\textwidth]{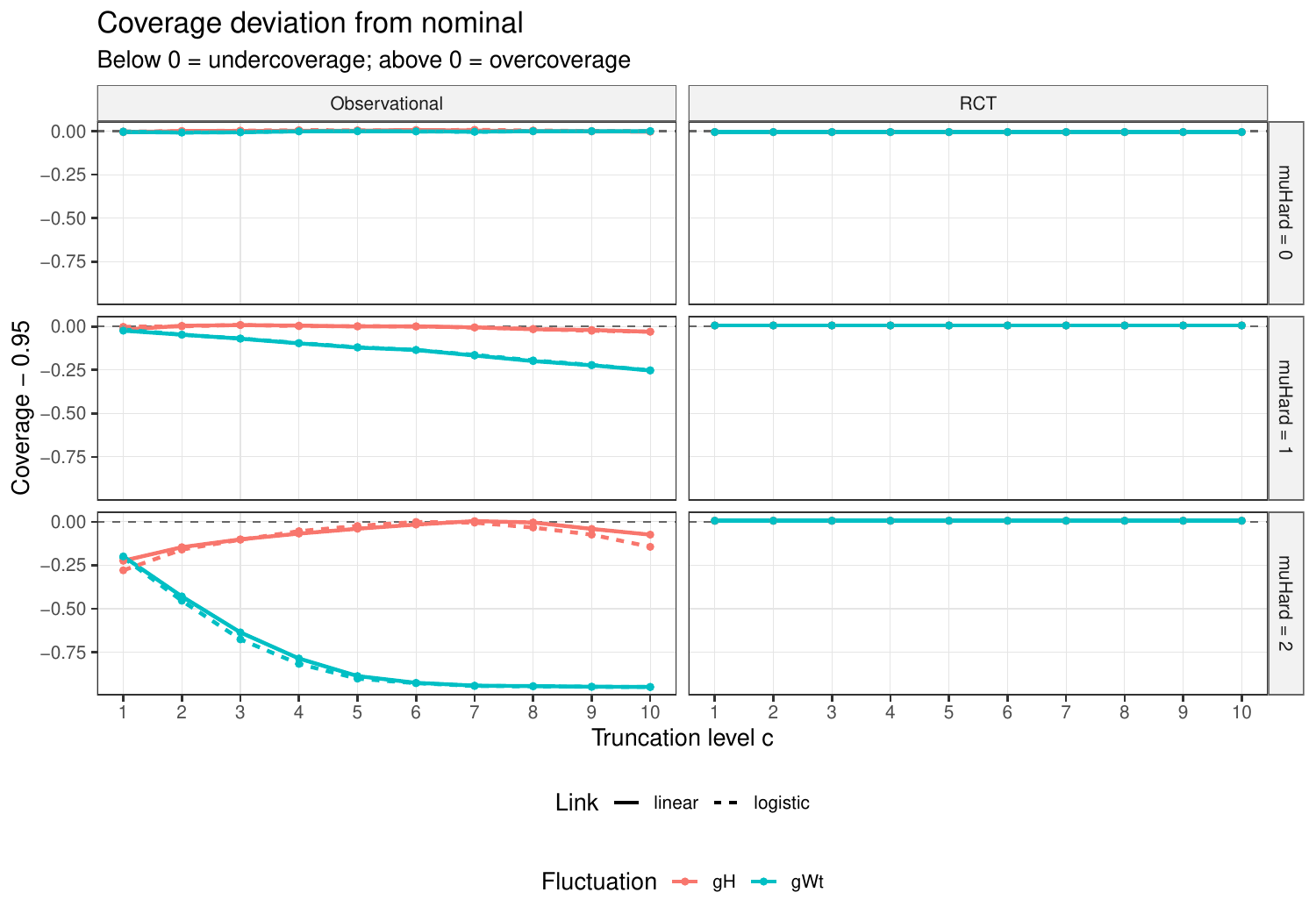}
    \caption{Coverage deviation from the nominal 95\% level across truncation levels $c=1,\dots,10$. Values below 0 indicate undercoverage and values above 0 indicate overcoverage. Curves compare fluctuation type ($gH$ vs.\ $gWt$) and link (logistic vs.\ linear), stratified by outcome hardness ($\mu_{\text{hard}}$) and study type (observational vs.\ RCT).}
    \label{fig:coverage-minus-nominal}
\end{figure}

This section provides detailed simulation results that support the main findings, together with additional simulation comparisons. \Cref{fig:n500_mu0pos1,fig:n1000_mu0pos1,fig:n500_mu0pos2,fig:n1000_mu0pos2,fig:n500_mu0pos3,fig:n1000_mu0pos3,fig:n500_mu1pos1,fig:n1000_mu1pos1,fig:n500_mu1pos2,fig:n1000_mu1pos2,fig:n500_mu1pos3,fig:n1000_mu1pos3,fig:n500_mu2pos1,fig:n1000_mu2pos1,fig:n500_mu2pos2,fig:n1000_mu2pos2,fig:n500_mu2pos3} report the full set of detailed metrics for the adaptive truncation procedures across scenarios. \Cref{fig:mse-ratio-link} compares logistic and linear links: under the RCT setting, logistic and linear links give essentially the same MSE, and in the observational setting the same is true for \texttt{gWt}, whereas for \texttt{gH} we observe more noticeable differences under higher outcome-model misspecification. \Cref{fig:mse-ratio-fluct} compares \texttt{gWt} and \texttt{gH}: under the RCT setting they are very similar across misspecification levels, while under observational settings with larger misspecification, \texttt{gH} is generally preferred. Coverage comparisons contrasting link function and targeting strategy are shown in \Cref{fig:coverage-minus-nominal}. \Cref{tab:mse_appendix} shows that in the absence of practical positivity violations, all methods perform similarly. Finally, \Cref{tab:variance_methods_k3_high} reports the variance estimates based on Monte Carlo variability, the sample variance of the EIF, the plug-in estimator, and the targeted bootstrap; the same qualitative observation as in \Cref{fig:relative-var-gH} holds, and the \texttt{gWt} case shows a similar pattern.

\begin{table}[htbp]
\centering
\caption{Monte Carlo MSE by truncation level $c$ under scenarios with and without practical positivity violations.}
\label{tab:mse_appendix}
\resizebox{\textwidth}{!}{%
\begin{tabular}{c|ccccc|ccccc}
\hline
& \multicolumn{5}{c|}{No Positivity Violation} & \multicolumn{5}{c}{Positivity Violation} \\
\cline{2-11}
$c$ 
& init & log gH & log gWt & linear gH & linear gWt 
& init & log gH & log gWt & linear gH & linear gWt \\
\hline
1  & 0.008 & 0.003 & 0.003 & 0.003 & 0.003 & 2.364 & 1.306 & 0.280 & 6.312 & 0.259 \\
2  & 0.008 & 0.003 & 0.003 & 0.003 & 0.003 & 2.364 & 0.706 & 0.301 & 1.452 & 0.277 \\
3  & 0.008 & 0.003 & 0.003 & 0.003 & 0.003 & 2.364 & 0.347 & 0.341 & 0.524 & 0.312 \\
4  & 0.008 & 0.003 & 0.003 & 0.003 & 0.003 & 2.364 & 0.179 & 0.381 & 0.238 & 0.349 \\
5  & 0.008 & 0.003 & 0.003 & 0.003 & 0.003 & 2.364 & 0.101 & 0.420 & 0.124 & 0.385 \\
6  & 0.008 & 0.003 & 0.003 & 0.003 & 0.003 & 2.364 & 0.066 & 0.456 & 0.074 & 0.418 \\
7  & 0.008 & 0.003 & 0.003 & 0.003 & 0.003 & 2.364 & 0.053 & 0.490 & 0.051 & 0.450 \\
8  & 0.008 & 0.003 & 0.003 & 0.003 & 0.003 & 2.364 & 0.052 & 0.521 & 0.043 & 0.479 \\
9  & 0.008 & 0.003 & 0.003 & 0.003 & 0.003 & 2.364 & 0.059 & 0.551 & 0.043 & 0.507 \\
10 & 0.008 & 0.003 & 0.003 & 0.003 & 0.003 & 2.364 & 0.070 & 0.579 & 0.048 & 0.534 \\
\hline
\end{tabular}%
}
\end{table}

\begin{table}[t]
\centering
\caption{Variance estimates across truncation levels for the clever-covariate–scaled targeting (gH) and loss-weighted targeting (gWt) estimators in the setting with high outcome misspecification and severe positivity violations ($\kappa_{\mathrm{pos}}=3$, $n=1000$). MC denotes the Monte Carlo variance, EIF the efficient influence function variance estimator, Plug-in the plug-in estimator, and TB the targeted bootstrap estimator.}
\label{tab:variance_methods_k3_high}
\scriptsize
\begin{tabular}{c|cccc|cccc}
\toprule
& \multicolumn{4}{c|}{gH} & \multicolumn{4}{c}{gWt} \\
\cline{2-9}
$c$
& MC & EIF & Plug-in & TB
& MC & EIF & Plug-in & TB \\
\midrule
1  & 0.8463 & 0.0563 & 0.1355 & 0.4747
   & 0.1123 & 0.0914 & 0.1275 & 0.1021 \\
2  & 0.6237 & 0.0382 & 0.0774 & 0.4552
   & 0.0630 & 0.0529 & 0.0726 & 0.0662 \\
3  & 0.3674 & 0.0261 & 0.0544 & 0.3448
   & 0.0451 & 0.0371 & 0.0513 & 0.0524 \\
4  & 0.2229 & 0.0193 & 0.0420 & 0.2536
   & 0.0361 & 0.0289 & 0.0399 & 0.0430 \\
5  & 0.1471 & 0.0153 & 0.0342 & 0.1993
   & 0.0304 & 0.0238 & 0.0327 & 0.0370 \\
6  & 0.1060 & 0.0126 & 0.0289 & 0.1569
   & 0.0265 & 0.0203 & 0.0277 & 0.0327 \\
7  & 0.0813 & 0.0108 & 0.0251 & 0.1312
   & 0.0237 & 0.0178 & 0.0241 & 0.0292 \\
8  & 0.0651 & 0.0096 & 0.0222 & 0.1117
   & 0.0217 & 0.0158 & 0.0214 & 0.0267 \\
9  & 0.0540 & 0.0086 & 0.0200 & 0.0975
   & 0.0202 & 0.0143 & 0.0193 & 0.0248 \\
10 & 0.0463 & 0.0079 & 0.0181 & 0.0860
   & 0.0191 & 0.0131 & 0.0176 & 0.0233 \\
\bottomrule
\end{tabular}
\end{table}

\end{document}